%% file: paper.tex
\documentclass[11pt]{article}

\usepackage[margin=1in]{geometry}
\usepackage{amsmath,amssymb,amsfonts}
\usepackage{graphicx}
\usepackage{hyperref}
\usepackage[T1]{fontenc}
\usepackage{times}  % Times font for better readability

\usepackage{bbm}
\usepackage{booktabs}   %% For formal tables
\usepackage{subcaption} %% For complex figures with subfigures/subcaptions
\usepackage{bussproofs}
\usepackage[cal=boondoxo]{mathalfa}
\DeclareMathAlphabet{\mathpzc}{OT1}{pzc}{m}{it}

\newtheorem{theorem}{Theorem}[section]

\newtheorem{definition}{Definition}
\newtheorem{corollary}{Corollary}

\usepackage{color}
\usepackage{xspace}

\input{macros}

\usepackage{bm}

\input{preamble}
\usepackage{pifont}% http://ctan.org/pkg/pifont
\usepackage{multicol,tabularx,capt-of}
\usepackage{multirow}

\title{Quantum Simulation Programming via Typing}

\author{
Liyi Li\\
Iowa State University\\
\texttt{liyili2@iastate.edu}
\and
Federico Zahariev\\
Iowa State University\\
\texttt{fzahari@iastate.edu}
\and
Chandeepa Dissanayake\\
Iowa State University\\
\texttt{chandeep@iastate.edu}
\and
Jae Swanepoel\\
Iowa State University\\
\texttt{jaeswan@iastate.edu}
\and
Amr Sabry\\
Indiana University\\
\texttt{sabry@iu.edu}
\and
Mark S. Gordon\\
Iowa State University\\
\texttt{mgordon@iastate.edu}
}

\date{\today}

\begin{document}

\maketitle

%% Abstract
\begin{abstract}
\input{abstract}

\end{abstract}

%% Keywords (optional for arXiv)
\textbf{Keywords:} quantum computing, programming languages, type systems, quantum simulation

\input{intros}                   % intro
\input{overview}                % overview
\input{semantics}                   % formal developement
\input{compilation}

\input{case}

\input{comparison}
\input{conclusion}

\section*{Acknowledgments}
This material is based upon work supported by the National Science Foundation under Grant No. OSI-2435255.

%% Bibliography
\bibliographystyle{plain}  % Changed from eptcs to plain for arXiv
\bibliography{paper}

\newpage
\appendix
\input{background}

\input{appendix}

\end{document}

%% file: macros.tex
\usepackage{listings}
\newcommand{\ignore}[1]{{}}

\newcommand{\syntaxDef}[3]{\rulebox{%
\syntaxKeyword$#1\mathrel{::=}{#2}$ \ifthenelse{\equal{#3}{}}{}{[#3]}%
}%
}

\makeatletter
\newcommand{\shorteq}{%
  \settowidth{\@tempdima}{-}% Width of hyphen
  \resizebox{\@tempdima}{\height}{=}%
}
\makeatother

%% file: preamble.tex
\usepackage{xcolor}
\usepackage{colortbl}

\usepackage[utf8]{inputenc}

\usepackage{amsmath}
\usepackage{stmaryrd}
\usepackage{hyperref}

\usepackage{subcaption}
\usepackage{wrapfig}

\usepackage{xspace}
\usepackage{float}
\usepackage{balance}

\usepackage{textcomp} %just for a vertical quote

\usepackage{changepage}
\usepackage{array,etoolbox}

\preto\tabular{\setcounter{magicrownumbers}{0}}
\newcounter{magicrownumbers}

\usepackage{cleveref} % must be loaded after hyperref and amsmath

\usepackage{wrapfig}
\usepackage{caption}

\usepackage{longtable}
\usepackage{tabu}

\usepackage{wasysym}
\usepackage{mathpartir}
\usepackage{mathtools} %psmallmatrix
\usepackage{xfrac} % fractions

\usepackage[qm,braket]{qcircuit}

\newcommand{\qass}[2]{{#1}\;{\leftarrow}\;{#2}}

\newcommand{\qif}[2]{\texttt{{if}\,(}{#1}\texttt{)\,\{}{#2}\texttt{\}}}

\colorlet{kwd}{black!80!green}
\definecolor{spec1}{RGB}{78, 131, 162}
\definecolor{spec0}{RGB}{66, 102, 136}
\definecolor{lespec}{RGB}{30, 80, 180}
\colorlet{spec}{lespec}
\colorlet{auto}{lespec!35!lightgray}
\colorlet{stack}{magenta}

\newcommand{\qafny}{\rulelab{Qafny}\xspace}

\newcommand{\sqir}{SQIR\xspace}

\newcommand{\qwire}{{Qwire}\xspace}
\newcommand{\qbricks}{{QBricks}\xspace}

\newcommand{\voqc}{\textsc{voqc}\xspace}

\newcommand{\myparagraph}[1]{\noindent\paragraph{\textbf{#1}}}

\usepackage{tikz}

\newcommand{\slen}[1]{|#1|}
\newcommand{\CC}{\ensuremath{\mathbb{C}}\xspace}
\usepackage{pgf}

\usepackage{tikz} % circuit diagrams
\usetikzlibrary{%
  arrows,%
  shapes.misc,% wg. rounded rectangle
  shapes.geometric, % diamonds
  shapes.arrows,%
  shapes.callouts,
  shapes.gates.logic.US,
  chains,%
  matrix,%
  positioning,% wg. " of "
  scopes,%
  decorations.pathmorphing,% /pgf/decoration/random steps | erste Graphik
  decorations.text,
  decorations.pathreplacing, % braces
  shadows,%
  automata,
  fit, calc, arrows.meta
}

\tikzset{ machine/.style={
    rectangle,
    minimum width=25mm,
    minimum height=18mm,
    text width=24mm,
    align=center,
    very thick,
    draw=black,
    color=black,
    fill=white,
  }
}

\DeclarePairedDelimiter\abs{\lvert}{\rvert}
\DeclarePairedDelimiter\norm{\lVert}{\rVert}

\makeatletter
\let\oldabs\abs
\def\abs{\@ifstar{\oldabs}{\oldabs*}}
\let\oldnorm\norm
\def\norm{\@ifstar{\oldnorm}{\oldnorm*}}
\makeatother

\makeatletter
\DeclareRobustCommand{\vardivision}{%
  \mathbin{\mathpalette\@vardivision\relax}% 
}
\newcommand{\@vardivision}[2]{%
  \reflectbox{$\m@th\smallsetminus$}%
}
\makeatother

\newcommand{\ie}{\emph{i.e.,}\xspace}

\usepackage{booktabs}

\usepackage{alltt}
\usepackage{listings,lstcoq}
\definecolor{ltblue}{rgb}{0,0.4,0.4}
\definecolor{dkblue}{rgb}{0,0.1,0.6}
\definecolor{dkgreen}{rgb}{0,0.35,0}
\definecolor{dkviolet}{rgb}{0.3,0,0.5}
\definecolor{dkred}{rgb}{0.5,0,0}
\usepackage[export]{adjustbox}

\newcommand{\code}[1]{{\small\texttt{#1}}}
 \newcommand{\rulelab}[1]{{\small \textsc{#1}}}

\newcommand{\sch}[3]{\Msum_{j=0}^{#1}{#2 {#3}}}

\DeclareMathOperator*{\Motimes}{\text{\raisebox{0.25ex}{\scalebox{0.8}{$\bigotimes$}}}}
\DeclareMathOperator*{\Msum}{\text{\raisebox{0.25ex}{\scalebox{0.8}{$\sum$}}}}

\newcommand{\lambdae}[3]{\lambda #1 : #2 .\, #3}
\newcommand{\tjudge}[3]{#1 \vdash #2 \triangleright #3}

\newcommand{\teq}[2]{ #1 \equiv #2}

\newcommand{\sapp}[2]{#1\;#2}

\newcommand{\zero}{\mathpzc{O}}

\newcommand{\dualb}[3]{#1\ket{#2}_{0}\ket{#3}_{1}}

\newcommand{\sdag}[1]{#1^{\dag}}

\newcommand{\pmx}{\cn{p}}
\newcommand{\hmx}{\cn{h}}

\newcommand{\sminus}{\texttt{-}}
\newcommand{\splus}{\texttt{+}}
\newcommand{\eexp}[1]{\cn{exp}(\sminus i #1)}
\newcommand{\quan}[3]{\mathpzc{#1}^{#2}(#3)}

\newcommand{\funsa}[3]{\mathpzc{#1}(#2)(#3)}
\newcommand{\elog}[1]{\sminus i\cn{log}(#1)}

\newcommand{\qsnd}{$\textsc{QBlue}$\xspace}

\newcommand{\cn}[1]{\texttt{#1}}

\newcommand{\denote}[1]{\llbracket #1 \rrbracket\xspace}
\newcommand{\dabs}[1]{|\!| #1 |\!|}

\newcommand{\braketa}[2]{\ensuremath{\langle #1 | #2 \rangle}\xspace} % defined as ip in qcircuit

\usepackage{newunicodechar}
\let\Alpha=A
\let\Beta=B
\let\Epsilon=E
\let\Zeta=Z
\let\Eta=H
\let\Iota=I
\let\Kappa=K
\let\Mu=M
\let\Nu=N
\let\Omicron=O
\let\omicron=o
\let\Rho=P
\let\Tau=T
\let\Chi=X

\newunicodechar{Α}{\ensuremath{\Alpha}}
\newunicodechar{α}{\ensuremath{\alpha}}
\newunicodechar{Β}{\ensuremath{\Beta}}
\newunicodechar{β}{\ensuremath{\beta}}
\newunicodechar{Γ}{\ensuremath{\Gamma}}
\newunicodechar{γ}{\ensuremath{\gamma}}
\newunicodechar{Δ}{\ensuremath{\Delta}}
\newunicodechar{δ}{\ensuremath{\delta}}
\newunicodechar{Ε}{\ensuremath{\Epsilon}}
\newunicodechar{ε}{\ensuremath{\epsilon}}
\newunicodechar{ϵ}{\ensuremath{\varepsilon}}
\newunicodechar{Ζ}{\ensuremath{\Zeta}}
\newunicodechar{ζ}{\ensuremath{\zeta}}
\newunicodechar{Η}{\ensuremath{\Eta}}
\newunicodechar{η}{\ensuremath{\eta}}
\newunicodechar{Θ}{\ensuremath{\Theta}}
\newunicodechar{θ}{\ensuremath{\theta}}
\newunicodechar{ϑ}{\ensuremath{\vartheta}}
\newunicodechar{Ι}{\ensuremath{\Iota}}
\newunicodechar{ι}{\ensuremath{\iota}}
\newunicodechar{Κ}{\ensuremath{\Kappa}}
\newunicodechar{κ}{\ensuremath{\kappa}}
\newunicodechar{Λ}{\ensuremath{\Lambda}}
\newunicodechar{λ}{\ensuremath{\lambda}}
\newunicodechar{Μ}{\ensuremath{\Mu}}
\newunicodechar{μ}{\ensuremath{\mu}}
\newunicodechar{Ν}{\ensuremath{\Nu}}
\newunicodechar{ν}{\ensuremath{\nu}}
\newunicodechar{Ξ}{\ensuremath{\Xi}}
\newunicodechar{ξ}{\ensuremath{\xi}}
\newunicodechar{Ο}{\ensuremath{\Omicron}}
\newunicodechar{ο}{\ensuremath{\omicron}}
\newunicodechar{Π}{\ensuremath{\Pi}}
\newunicodechar{π}{\ensuremath{\pi}}
\newunicodechar{ϖ}{\ensuremath{\varpi}}
\newunicodechar{Ρ}{\ensuremath{\Rho}}
\newunicodechar{ρ}{\ensuremath{\rho}}
\newunicodechar{ϱ}{\ensuremath{\varrho}}
\newunicodechar{Σ}{\ensuremath{\Sigma}}
\newunicodechar{σ}{\ensuremath{\sigma}}
\newunicodechar{ς}{\ensuremath{\varsigma}}
\newunicodechar{Τ}{\ensuremath{\Tau}}
\newunicodechar{τ}{\ensuremath{\tau}}
\newunicodechar{Υ}{\ensuremath{\Upsilon}}
\newunicodechar{υ}{\ensuremath{\upsilon}}
\newunicodechar{Φ}{\ensuremath{\Phi}}
\newunicodechar{φ}{\ensuremath{\phi}}
\newunicodechar{ϕ}{\ensuremath{\varphi}}
\newunicodechar{Χ}{\ensuremath{\Chi}}
\newunicodechar{χ}{\ensuremath{\chi}}
\newunicodechar{Ψ}{\ensuremath{\Psi}}
\newunicodechar{ψ}{\ensuremath{\psi}}
\newunicodechar{Ω}{\ensuremath{\Omega}}
\newunicodechar{ω}{\ensuremath{\omega}}

\newunicodechar{ℕ}{\ensuremath{\mathbb{N}}}
\newunicodechar{∅}{\ensuremath{\emptyset}}

\newunicodechar{∙}{\ensuremath{\bullet}}
\newunicodechar{≈}{\ensuremath{\approx}}
\newunicodechar{≅}{\ensuremath{\cong}}
\newunicodechar{≡}{\ensuremath{\equiv}}
\newunicodechar{≤}{\ensuremath{\le}}
\newunicodechar{≥}{\ensuremath{\ge}}
\newunicodechar{≠}{\ensuremath{\neq}}
\newunicodechar{∀}{\ensuremath{\forall}}
\newunicodechar{∃}{\ensuremath{\exists}}
\newunicodechar{±}{\ensuremath{\pm}}
\newunicodechar{∓}{\ensuremath{\pm}}
\newunicodechar{·}{\ensuremath{\cdot}}
\newunicodechar{⋯}{\ensuremath{\cdots}}
\newunicodechar{…}{\ensuremath{\ldots}}
\newunicodechar{∷}{~\mathrel{:\!\!\!:}~}
\newunicodechar{×}{\ensuremath{\times}}
\newunicodechar{∞}{\ensuremath{\infty}}
\newunicodechar{→}{\ensuremath{\to}}
\newunicodechar{←}{\ensuremath{\leftarrow}}
\newunicodechar{⇒}{\ensuremath{\Rightarrow}}
\newunicodechar{↦}{\ensuremath{\mapsto}}
\newunicodechar{↝}{\ensuremath{\leadsto}}
\newunicodechar{∨}{\ensuremath{\vee}}
\newunicodechar{∧}{\ensuremath{\wedge}}
\newunicodechar{⊢}{\ensuremath{\vdash}}
\newunicodechar{⊣}{\ensuremath{\dashv}}
\newunicodechar{∣}{\ensuremath{\mid}}
\newunicodechar{∈}{\ensuremath{\in}}
\newunicodechar{⊆}{\ensuremath{\subseteq}}
\newunicodechar{⊂}{\ensuremath{\subset}}
\newunicodechar{∪}{\ensuremath{\cup}}
\newunicodechar{⋓}{\ensuremath{\Cup}}
\newunicodechar{∉}{\ensuremath{\not\in}}
\newunicodechar{√}{\ensuremath{\sqrt}}

\newunicodechar{⊸}{\ensuremath{\multimap}}
\newunicodechar{⊗}{\ensuremath{\otimes}}
\newunicodechar{⨂}{\ensuremath{\bigotimes}}
\newunicodechar{⊕}{\ensuremath{\oplus}}
\newunicodechar{〈}{\ensuremath{\langle}}
\newunicodechar{⟨}{\ensuremath{\langle}}
\newunicodechar{⟩}{\ensuremath{\rangle}}
\newunicodechar{〉}{\ensuremath{\rangle}}
\newunicodechar{¡}{\ensuremath{\upsidedownbang}}
\newunicodechar{∘}{\ensuremath{\circ}}
\newunicodechar{†}{\ensuremath{\dagger}}
\newunicodechar{⊤}{\ensuremath{\top}}
\newunicodechar{⊥}{\ensuremath{\bot}}

\newunicodechar{〚}{\ensuremath{\llbracket}}
\newunicodechar{〛}{\ensuremath{\rrbracket}}

\usepackage{etoolbox} % replaces ifthen package
\newtoggle{comments}
\toggletrue{comments}
  \usepackage[normalem]{ulem}

\iftoggle{comments}{
  \newcommand{\fixme}[1]{\textbf{\textcolor{red}{[ Fixme: #1]}}}
  \newcommand{\todo}[1]{\textbf{\textcolor{green}{[ TODO: #1 ]}}}
  \newcommand{\mwh}[1]{\textbf{\textcolor{red}{[ Mike: #1 ]}}}
  
  \newcommand{\khh}[1]{\textbf{\textcolor{orange}{[ Kesha: #1 ]}}}
  \newcommand{\shh}[1]{\textbf{\textcolor{purple}{[ Shih-Han: #1 ]}}}
  \newcommand{\liyi}[1]{\textbf{\textcolor{orange}{[ Liyi: #1 ]}}}
  
 \newcommand{\chand}[1]{\textbf{\textcolor{blue}{[ Chandeepa: #1 ]}}}
  \newcommand{\oth}[2]{\textbf{\textcolor{red}{[ #1: #2 ]}}}
  \newcommand{\xwu}[1]{\textbf{\textcolor{purple}{[ Xiaodi: #1 ]}}}
  
  \newcommand{\ynote}[1]{\textbf{\textcolor{magenta}{[ Yi: #1 ]}}}

  \colorlet{MZ}{violet!80!pink}

  \newcommand{\mzr}[1]{{\color{MZ}{#1}}}
  \newcommand{\was}[1]{}
  
  \NewCommandCopy{\Creff}{\Cref}
  \renewcommand{\Cref}[1]{\mbox{\Creff{#1}}}

  \usepackage[inline]{enumitem}
  \colorlet{LC}{cyan!31!teal}

  \usepackage[inline]{enumitem}
}{
  \newcommand{\fixme}[1]{}
  \newcommand{\todo}[1]{}
  \newcommand{\rnr}[1]{}
  \newcommand{\mwh}[1]{}  
  \newcommand{\khh}[1]{}
  \newcommand{\liyi}[1]{}
  \newcommand{\shh}[1]{}
  \newcommand{\xwu}[1]{}
  \newcommand{\oth}[2]{}
  \newcommand{\mzr}[1]{}
  \newcommand{\chand}[1]{}

  \newcommand{\ynote}[1]{}
}

\newtoggle{submission}
\toggletrue{submission}
\iftoggle{submission}{
  
}{
  
}

%% file: abstract.tex
%Quantum simulation intends to analyze quantum particle system behaviors, and many current quantum computation compilation frameworks have been developed to perform quantum simulations on a quantum computer. These compilers solely focus on exploring the ability of digital and analog quantum computers by programming quantum particle systems in terms of Pauli strings or digital quantum circuits; such programs are challenging to program by quantum simulation users in physics, chemistry, and biology.
%We propose \qsnd, the first programming language to allow users to think of programs as second quantization based Hamiltonians for describing their desired quantum particle system behaviors.
%A novel type system for \qsnd is developed to crystalize different particle states in different particle systems, where we can view quantum computers as particle systems of a certain type. Using the type system, we permit the compilation of the quantum simulation of a \qsnd program to both digital and analog quantum computers.
%Based on \qsnd, users can program the desired quantum particle system properties and analyze them in a quantum computer.

Quantum simulations are designed to model quantum systems, and many compilation frameworks have been developed for executing such simulations on quantum computers. Most compilers leverage the capabilities of digital and analog quantum computers by representing quantum particle systems with Pauli strings or digital quantum circuits, making it challenging for users in physics, chemistry, and biology to program simulations effectively. \qsnd is proposed as the first programming language for describing the behaviors of quantum systems in terms of second quantization Hamiltonians. Within \qsnd, a novel type system is proposed to clearly define states across different quantum systems and treat quantum computers as quantum particle systems of specific types. The type system is compatible with the compilation of quantum simulations expressed in \qsnd for digital and analog quantum computers. With \qsnd, users can specify the desired quantum particle system and model the system on quantum computers.

%% file: intros.tex
\section{Introduction}\label{sec:intro}

%Quantum computing has led to significant milestones~\cite{Bravyi2024,king2024computational}, bringing new capacity to solve computationally difficult problems that are intractable using classical computers, e.g., Shor's algorithm~\cite{shors} can factor a number in polynomial time, which is not known to be polynomial-time-computable in the classical setting.
%Many quantum programming languages have been developed, mainly based on two approaches, digital-based and analog-based, as shown in \Cref{fig:process}.
%The dominant approach builds on digital-based quantum circuits~\cite{VOQC,10.1145/3519939.3523433,oracleoopsla,ccx-adder,quilc,Fagan2018,reverC,scaffCCnew,Nam2018,quantumssa,ripple-carry-mod,qft-adder}. There are additionally approaches that build on analog quantum computing~\cite{Daley2022,10.1145/3632923}  by viewing a quantum computer as a physical system simulator.

%FZ: The above is OK. Below is just an alternative shorter wording of the same:

\begin{figure}[h]
\vspace*{-0.5em}
  \includegraphics[width=0.9\textwidth]{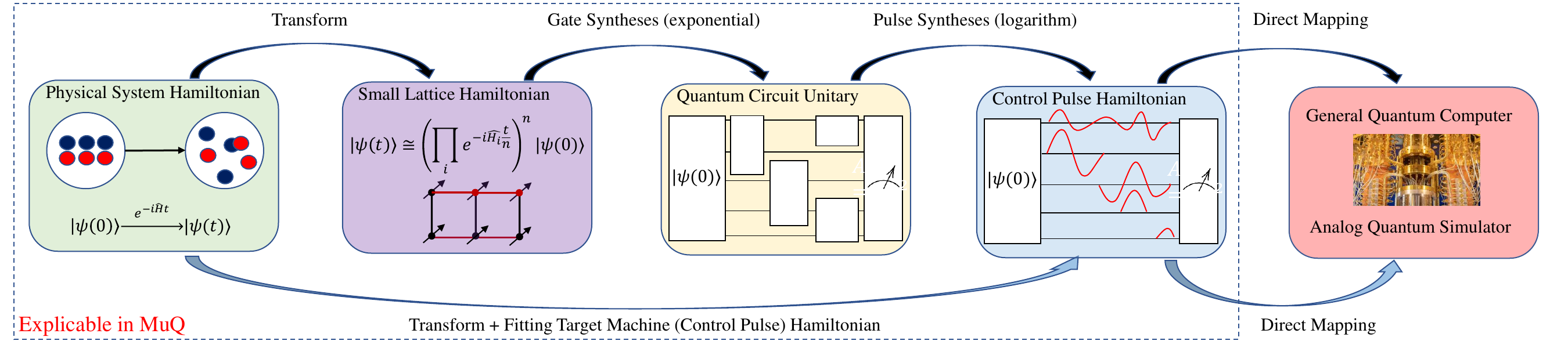}
  \vspace*{-0.5em}
     \caption{Digital-based (top black arrows) and analog-based (bottom blue arrows) compilation procedures. Notice that the target machines are the same for the two compilations but with different views.}\label{fig:process}
\vspace*{-0.5em}
\end{figure}

Quantum computing has made significant progress in recent years~\cite{Bravyi2024,king2024computational} to solve computational problems that are otherwise intractable for classical computers. For instance, Shor's algorithm~\cite{shors} can factor large numbers in polynomial time, a task that is not known to be solvable classically.
Quantum programming languages are categorized into two groups based on the two main approaches: digital and analog, as depicted in \Cref{fig:process}. The dominant approach is based on digital quantum circuits \cite{VOQC,10.1145/3519939.3523433,oracleoopsla,ccx-adder,quilc,Fagan2018,reverC,scaffCCnew,Nam2018,quantumssa,ripple-carry-mod,qft-adder}, while analog quantum computing \cite{Daley2022,10.1145/3632923} operates as if the quantum computer is a simulator of physical systems.
%Yet all these developments only scratch the surface of a much richer quantum landscape. 

%FZ: The above is OK. Below is just an alternative shorter wording of the same:

The flow in Fig. \Cref{fig:process} shows the two approaches and their compilation strategies.
To describe a physical system in analog quantum computing, users typically describe the energy flow of quantum particles in the system as Hamiltonians (Hermitian matrices).
Such Hamiltonian-based computational approaches for quantum particles are already well used by experimentalists, physicists, chemists, and other domain scientists, with or without quantum hardware, to design new materials~\cite{Emani2021,Fedorov2021,wcms.1481}, study chemical processes~\cite{Lanyon2010,Cao2019,Evangelista2023,RevModPhys.92.015003}, and to simulate many body microscopic systems~\cite{Zhang_2021,osti_1782924,du2023multinucleon}. 
These have led to a rich trove of computational abstractions that are underutilized in creating quantum programming languages.

%FZ: Below is a shorter and much better (!!!) wording:
Unfortunately, no quantum computer language is based on programming quantum particle system properties, e.g., Hamiltonian simulations.
Existing Hamiltonian simulation frameworks are based on Pauli strings \cite{10.1145/3632923,10.1145/3503222.3507715,10.1145/3470496.3527394}, describing what the quantum computer is capable of, not what the programmers intend to do.
We propose \qsnd, the first programming language for programming quantum particle system computation, based on second quantization, a well-known physical quantization method for quantum particle systems.
At its core, \qsnd is a quantum particle system description language based on the idea of lattice-based Hamiltonian systems where the computation is presented as lattice graphs occupied by quantum particles.

In \qsnd, users can program particle system behaviors in terms of second quantization operations, such as creators and annihilators.
\qsnd also includes a type system to distinguish different particles.
In a quantum computer, the elementary state entity is a single qubit.
However, this becomes insufficient to analyze quantum particle systems because they might have many different types of particles, such as bosons and fermions.
When writing a second quantization operation, users might not immediately be aware of the types of particles the operation deals with.
In \qsnd, we develop a type system to classify particles occupying different sites in a lattice graph, enabling the description of the interactions of various particles.
Via our typed-guided language, we can compile a \qsnd program to a quantum computer based on a qubit state system.
Our compiler is general enough to compile to both a digital and analog quantum computer.

%the below one can be merged with the above one.
\qsnd is intended to solve these problems by introducing a unified formalism for describing quantum particle system Hamiltonians as a programming language. It also has a type system strong enough to distinguish between types of particles and to declare what operations can be performed on each, making the code easier to understand, maintain, and reuse in quantum simulation. The paper's contributions are listed as follows.

\begin{itemize}

\item We define the \qsnd language syntax and semantics in \Cref{sec:overview} and \Cref{sec:formal}. %and show the key advantages in \Cref{sec:compilation}.

\item In \Cref{sec:formal}, we develop a type system for \qsnd that is general enough to describe many kinds of quantum lattice models, including qubit-based quantum computing systems. We prove the type soundness of \qsnd in Coq, which guarantees that any \qsnd program, when typed as Hermitian, can be safely translated to quantum computers.

\item In \Cref{sec:qcompile}, we show a compilation procedure to compile \qsnd to both digital- and analog-based quantum circuits. Furthermore, we explain how this generalization leads to improved optimization by considering the hardware specificities.

\item In \Cref{sec:boson}, we show an example usage of \qsnd for compiling bosonic systems. We demonstrate a case using \qsnd to analyze fermionic systems in \Cref{sec:fermions}.

\end{itemize}

\ignore{
Previously, the former world \cite{vandenBerg2020circuitoptimization,Smith2019a,Berry_2015b,10.5555/2481569.2481570} typically shows that the ability of implementing Hamiltonian simulation via quantum computer gates without showing the meaning of the gate combination,
while the latter world (the works above) represents problems in terms of Hamiltonians and provided some computation on the Hamiltonians, without indicating if the computation can be performed in a quantum computer. \qsnd provides the platform for the two worlds to communicate together.
Especially, the linking between \qsnd and $\lambda$-calculus permits authors to understand quantum systems in terms of intuitive functional language,
such that Hamiltonian operations can be understood as function applications.
By doing so, we also lift the quantum algorithm finding, previously restricted to quantum circuit computing, to a broader high-performance computing (HPC) field.
There is a popular critization \cite{JackKrupansky,MattSwayne,RichardWaters} that quantum computers lack of applications.
The designs of new quantum algorithms should be in the lifted field instead of restricting to quantum computing fields,
since many algorithms showing quantum advantage act as subroutines of a large application.
For example, Energy computation in quantum chemistry \cite{Lee2023} might not have quantum advantage in executing it in a quantum computer,
but some subroutine computation might experience quantum advantage \cite{BLOMBERG2006969,Kovyrshin2023}.

Regarding the user perspective, the users that desperately need the quantum computing power might be from scientific computing fields and try to solve hard problems in physics \cite{Zhang_2021,osti_1782924,du2023multinucleon}, chemistry \cite{Lanyon2010,Cao2019,Evangelista2023,RevModPhys.92.015003}, and computational biology \cite{Emani2021,Fedorov2021,wcms.1481};
such problems are typically defined in terms of Hamiltonians with applications on these Hamiltonians, such as performing ground energy state computation and Hamiltonian simulation.
If both the user and the underlying machine levels of quantum computing are performing applications related to Hamiltonians, it indicates strongly the needs of a new Hamiltonian-based intermediate representation language for expressing quantum computing algorithms.

1. machine reality ---> quantum computers are simulators. analog quantum computing ---> directly implement problems in quantum computers.

2. user prosepctive, many applications are scintific comptuing problems, such as ....

3. if users and machines are both hamiltonians, why using unitary in the middle? % might not want to say this.

4. provide a chance to examine the necciecity of linking the two.
Whether or not we should use Hamiltonian or unitary, it is time to develop a user (programmer) friendly framework to provide a generalized view of quantum computing in the context of quantum mechanics.

reasons. 

a. time to connect users and machines. Many current works only provide one or two example cases, tend to merge everything together by mapping an example system in quantum circuits and states, probably with optimizations to make a case. Then, different way of mapping for optimization causes confusion on learning about original system, and how to compile systems to quantum computers in general, e.g., the descriptions of the physical systems are misunderstood due to the understand of the compiled version (CITE).

realize what quantum computers actually are t(1) type bosonic system.

b. correct optimizations in the machine and user context. current optimizations have no context, with the user and machine level details. 
enabling the analysis of viewing quantum computers as simulators, bypassing the circuit models.

c. provide typing for long existing problems in describing physical systems. fork states are not good for describing systems. a formalism of second quantizations. differet verions discring different systems, without having the context hard to understand, so we need a type system.

Having a functional computation model for second quantization, high-order functions for including both fermions and bosons. context switching between snapshot and simulation mode.

There are two observations based on machine reality and user prospective.
In the quantum machine level, each elementary quantum gate is implemented as controlled pulses, which is equivalent to perform a Hamiltonian simulation, shown in \Cref{fig:process}.
Essentially, every quantum computer has a fixed form of recognizable Hamiltonian.
To implement a quantum gate, we perform an eigendecomposition of the gate, construct the gate Hamiltonian that is recognizable by the quantum computer,
and simulate a certain time on the Hamiltonian as the simulation of the gate in the quantum computer.
This procedure represents the controlled pulse level programming in a quantum computer and it unveils that a quantum computer, in its machine level, is a Hamiltonian simulator that simulates gate behaviors.

}

%There are several reasons for the \qsnd development.
%The primary reason is to connect the quantum computing and the user worlds.
 
\ignore{
In quantum field theory, it is known as canonical quantization, in which the fields (typically as the wave functions of matter) are thought of as field operators, in a manner similar to how the physical quantities (position, momentum, etc.) are thought of as operators in first quantization. The key ideas of this method were introduced in 1927 by Paul Dirac \cite{Dirac1988}, and were later developed, most notably, by Pascual Jordan \cite{Jordan1928} and Vladimir Fock \cite{Fock1932,Reed:1975uy}. In this approach, the quantum many-body states are represented in the Fock state basis, which are constructed by filling up each single-particle state with a certain number of identical particles \cite{Becchi:2010}. The second quantization formalism introduces the creation and annihilation operators to construct and handle the Fock states, providing useful tools to the study of the quantum many-body theory.

We propose a 

Second quantization, also referred to as occupation number representation, is a formalism used to describe and analyze quantum many-body systems. In quantum field theory, it is known as canonical quantization, in which the fields (typically as the wave functions of matter) are thought of as field operators, in a manner similar to how the physical quantities (position, momentum, etc.) are thought of as operators in first quantization. The key ideas of this method were introduced in 1927 by Paul Dirac \cite{Dirac1988}, and were later developed, most notably, by Pascual Jordan \cite{Jordan1928} and Vladimir Fock \cite{Fock1932,Reed:1975uy}. In this approach, the quantum many-body states are represented in the Fock state basis, which are constructed by filling up each single-particle state with a certain number of identical particles \cite{Becchi:2010}. The second quantization formalism introduces the creation and annihilation operators to construct and handle the Fock states, providing useful tools to the study of the quantum many-body theory.

In this paper, we develop a programming language for second quantization, \qsnd, based on the combination of the second qauntization formalism \cite{Dirac1988} and simple typed $\lambda$-calculus \cite{Church1940,Church1956}.

Developing more and more comprehensive quantum programs and algorithms is essential for the continued practical development of quantum computing \cite{JackKrupansky,MattSwayne}.
Unfortunately, because quantum systems are inherently probabilistic and also must obey laws of quantum physics, traditional validation techniques based on run-time testing are virtually impossible to develop for large quantum algorithms,
This leaves \emph{formal methods} as a viable alternative for program checking, and yet these typically require a great effort; for example,
four experienced researchers needed two years to formally verify Shor's algorithm \cite{shorsprove}.

To alleviate the effort required for formal verification, many frameworks have been proposed to verify quantum algorithms \cite{qhoreusage,qhoare,qbricks,qsepa,qseplocal,VOQC}
using interactive theorem provers, such as Isabelle, Coq, and Why3, by building quantum semantic interpretations and libraries.
Some attempts towards proof automation have been made by creating new proof systems for quantum data structures such as Hilbert spaces; however,
building and verifying quantum algorithms in these frameworks are still time-consuming and require great human effort.
Meanwhile, automated verification is an active research field in classical computation with many proposed frameworks
\cite{HoareLogic,separationlogic,nat-proof-fun,nat-proof,nat-proof-frame,10.1145/3453483.3454087,arxiv.1609.00919,martioliet00rewriting,rosu-stefanescu-2011-nier-icse,rosu-stefanescu-ciobaca-moore-2013-lics,10.1007/978-3-642-17511-4_20,10.1007/978-3-642-03359-9_2}
showing strong results in reducing programmer effort when verifying classical programs. 
%Can classical automated verification frameworks be utilized for verifying quantum programs?
None of the existing quantum verification frameworks utilize these classical verification infrastructures, however.

Second quantization, also referred to as occupation number representation, is a formalism used to describe and analyze quantum many-body systems. In quantum field theory, it is known as canonical quantization, in which the fields (typically as the wave functions of matter) are thought of as field operators, in a manner similar to how the physical quantities (position, momentum, etc.) are thought of as operators in first quantization. The key ideas of this method were introduced in 1927 by Paul Dirac \cite{Dirac1988}, and were later developed, most notably, by Pascual Jordan \cite{Jordan1928} and Vladimir Fock \cite{Fock1932,Reed:1975uy}. In this approach, the quantum many-body states are represented in the Fock state basis, which are constructed by filling up each single-particle state with a certain number of identical particles \cite{Becchi:2010}. The second quantization formalism introduces the creation and annihilation operators to construct and handle the Fock states, providing useful tools to the study of the quantum many-body theory.

In this paper, we develop a programming language for second quantization, \qsnd, based on the combination of the second qauntization formalism \cite{Dirac1988} and simple typed $\lambda$-calculus \cite{Church1940,Church1956}.

\liyi{1. Quantum Computation limitation. 2. describe the situation: 1) the main customers of quantum computers are scientific computing, 2) machine is implemented in terms of Ham simulation (see overview, describe a little). --> why we ever need quantum circuit? 3. main point is to create a language for scientific computing (marry the world of physics and CS). 4. (maybe not be good) Remove the circuit level and create a language for directly linking customers and mechine basis. }

\liyi{Describe what Ham can do. How physics view the world --> snapshot of Ham, and unitary for simulation for movies. }
}

%% file: overview.tex
%\vspace*{-0.3em}
\section{Overview}\label{sec:overview}
%\vspace*{-0.3em}

%This section provides an overview for \qsnd based on boson-like particle systems \footnote{The boson-like system does not deal with anti-commutation as the fermion system requires, and we model the boson system with upper bounds on basis states.} (we encode a different fermion particle system for \qsnd in \Cref{sec:fermions}).

This section provides an overview of \qsnd based on boson systems\footnote{A boson system uses commutation rules. We model it with an upper bound on the number of basis vector states.}, with a fermion system for \qsnd is encoded in \Cref{sec:fermions}.
The quantum computation and second quantization background can be found in \Cref{sec:background}.

\begin{figure}[t]
{\small
  \[
  \begin{array}{c}
  \begin{array}{ll@{\;\;}c@{\;\;}l @{\qquad} ll@{\;\;}c@{\;\;}l @{\qquad} ll@{\qquad}l}
  \text{Nat. Num} & m, n, j, k & \in & \mathbb{N}
&
\text{Complex Number} & z & \in & {\mathbb{C}}
&
      \text{Variable} & x,y,f,g 
\end{array}
\\[0.3em]
  \begin{array}{l@{\quad}l@{\;\;}c@{\;\;}l@{}c@{\;\;\;}l} 
\text{Single Particle Basis Vector With }m\text{ States} & \eta & ::= & {\displaystyle \ket{k}} && k\in[0,m)
\\
n\text{ Particle Ket State} & w & ::= &  {\displaystyle  z \cdot \Motimes_{k=0}^{n-1} \eta_{k} }
%\\[0.2em]
%\text{Single Particle Type With }$n$\text{ Spins} & t(n,m) & ::= &  {\bigotimes^{n}\textcolor{spec}{\mathpzc{H}^{m}}}
%\\[1em]
%\text{Single Particle State} & v & ::= & {\displaystyle{\sum_j} z_j\eta_j }
\\[0.5em]
\text{Quantum State Type} & \iota & ::= & t(m) &\mid& \iota \otimes \iota
\\[0.2em]
\text{Quantum State} & \varphi & ::= & {\displaystyle {\sum_j} w_j  } &\mid& \zero
    \end{array}
        \end{array}
  \]
}
\vspace*{-1.2em}
  \caption{\qsnd state structure for particles. A $t(m)$ typed basis vector $\ket{j}$ can have $j \in [0,m)$.  We abbreviate $z \cdot \Motimes_{k=0}^{1} \eta_{k} \equiv z\, \eta_0$, and $z_1\eta_1 \otimes z_2\eta_2\equiv (z_1 * z_2)\eta_1 \eta_2 $.}
  \label{fig:data}
  \vspace*{-1em}
\end{figure}

\myparagraph{Quantum States.}\label{sec:states}
Traditionally, a quantum state in quantum computation is composed of qubits, i.e., each single site \footnote{In thie paper, we name a single location in a lattice-graph for interacting particles a site, and we overload the site concept as a single location in a quantum state.} in the quantum state is a qubit --- a two-dimensional Hilbert space; discussed in \Cref{appx:state}. In \qsnd, we generalize qubits to quantum particles, using Dirac notations as our data-structure representations for quantum states in \Cref{fig:data}. The key difference is that a single site in \qsnd is not necessarily a qubit described by a two-dimensional Hilbert space. Instead, we generalize it to an $m$-dimensional Hilbert space of a quantum particle.

A basis-vector $\ket{k}$ of the quantum particle has the vector number $k$, ranging from $0$ to $m-1$.
We can compose many single-site quantum particles via tensor operations, and we name a pair of an amplitude $z$ and tensor product of quantum particles as a ket state $w$ in \Cref{fig:data}. A quantum state $\varphi$ is defined as a linear combination of kets.

The Hilbert space dimensionality can vary across different sites. To address this complication, we create state types $\iota = t(m)$, which specify the single-site quantum particle type, relating it to an $m$ dimensional Hilbert space. We separate particle types using the tensor type symbol as $\iota \otimes \iota$ (this paper classifies tensor types as type constructs in our system and tensor operations as Hamiltonian program operations). Tensor types act as site location separators between particles. 
The paper assumes a lattice graph structure exists for specifying site adjacencies without detailing how such a graph can be constructed.

\myparagraph{Program Syntax.}\label{sec:syntax}
\Cref{fig:syntax} shows \qsnd's program syntax based on second quantization.
Essentially, the program syntax defines a Hamiltonian to express the interactions among quantum particles residing on sites within a lattice graph structure.
After such a program is defined, users can perform Hamiltonian time evolution ($\eexp{\sapp{e}{t}}$) to simulate the system $e$ (we use variables $e$ and $\hat{H}$ to describe a Hamiltonian) with respect to time $t$, which generates a unitary matrix, executable in a quantum computer. Additionally, users might apply measurement operations to compute the energy, detailed in \Cref{sec:groundenergy}.

The operation $z{a}:\quan{F}{\zeta}{t(m)}$ is a single particle annihilator with amplitude $z$, meaning that an amplitude-adjusted annihilator applies to a single-particle state.
$\sdag{e}$ is a conjugate transpose operator. Specifically, $\sdag{(z\, a:\quan{F}{\zeta}{t(m)}}$, can be viewed as a syntactic sugar of $\sdag{z}\sdag{a}:\quan{F}{\zeta}{t(m)}$, defines a creator. We also have a tensor product operator $e \otimes e$ acting as a particle site separator, a linear sum $e+e$ (can be viewed as quantum choices in \Cref{appx:linearsum}), and sequence operation $\sapp{e}{e}$, i.e., a matrix multiplication operation.

We now introduce the operation type syntax. We associate each single particle operation with a type $\quan{F}{\zeta}{t(m)}$, meaning that the single particle operation applies to a single particle state with type $t(m)$. If we view $e$ as a Hamiltonian, the type construct $\quan{F}{\zeta}{t(m)}$ refers to a function type $t(m) \to t(m)$, i.e., a function dealing with a single particle state. In this sense, the sequence operation $\sapp{e}{e}$ is essentially a function application.
The flag $\zeta$ indicates the kind of matrix a program is constructing, an ordinary matrix ($\pmx$) or a hermitian one ($\hmx$).
Eventually, every quantum computer executable Hamiltonian program $e$ must be a Hermitian matrix ($\hmx$) so that performing a matrix exponent $\eexp{\sapp{e}{t}}$ creates a unitary matrix executable in a quantum computer, but neither creators ($\sdag{a}$) nor annihilators ($a$) are Hermitian. In \qsnd, we utilize a type system to guide users to define a correct Hermitian matrix (Hamiltonian).
We include tensor types $\quan{F}{\zeta}{\iota}$, with $\iota=\iota_1 \otimes \iota_2$, for separating particle sites (\Cref{fig:data}); such types represent the tensor types ($\otimes$) to connect the operations applied on different sites.

\begin{figure}[t]
\vspace*{-0.5em}
{\small
  \[\begin{array}{l@{\quad}l@{\;\;}c@{\;\;}l@{\;\;}} 
%\text{Single Particle Op} & \alpha & ::= & z a_{\sigma}
%\\
%\text{Arith Type} & \xi & ::= & \cn{nat} \mid \mathbb{R} \mid \mathbb{C} \mid \xi \to \xi
%\\
\text{Data Type Flag} & \zeta & ::= & \pmx \mid \hmx
\\[0.2em]
\text{Quantum Operation Type} & \tau & ::= & \quan{F}{\zeta}{\iota}
%\\
%\text{Type} & \omega & ::= & \xi \mid \tau \mid \omega \to \omega
\\[0.2em]
       \text{Hamiltonian Expression} & \hat{H}, e & ::= &z\,{a}:\quan{F}{\zeta}{t(m)} \mid \sdag{e} \mid e \otimes e \mid {e}+{e} \mid \sapp{e}{e}
    \end{array}
  \]
}
  \vspace*{-0.8em}
  \caption{\qsnd syntax. Hamiltonians can be named as $e$ or $\hat{H}$.}
  \label{fig:syntax}
    \vspace*{-1em}
\end{figure}

\myparagraph{An Illustrative Example: Hubbard Model (System).}\label{sec:runningexample}
The Hubbard model is a way of describing quantum particle behaviors in a lattice graph,
and it is used to describe the tunnelling of quantum particles in one site to the other.
The Hubbard model can have many forms, for both fermions and bosons (Bose-Hubbard system \cite{peng2023quantum}),
and we list the common term below to describe the kinetic term for particle tunnelling, and $z_t$ is the of kinetic energy of electron tunnelling.
Here, we mainly focus on the programming aspect and the Hubbard model detail is in \Cref{sec:hubbard}.

{\small
\[
\hat{H} = z_t \sum_{j} \sapp{\sdag{a}(j)}{ a(j\splus 1)} + \sapp{\sdag{a}(j\splus 1)}{ a(j)}
\tag{3.2.2}\label{eq:hubbard}
\]
}

For an $n$ particle system, an index of an operator such as $j$ in $\sdag(a)(j)$ is a syntactic sugar for the tensor product of the identity operator $I$ repeated $n-1$ times, i.e., we create a length $n$ tensor product of $I$ in any position other than the $j$-th one and $\sdag(a)(j)$. For instance, the term above of the two-particle quantum Hubbard model Hamiltonian can be written as:

{\small
\[
\label{equ1}
\sapp{(\sdag{a} \otimes I)}{(I \otimes a)} + \sapp{(I \otimes \sdag{a})}{(a \otimes I)} + \sapp{(\sdag{a} \otimes I)}{(I \otimes a)} + \sapp{(I \otimes \sdag{a})}{(a \otimes I)}
\]
}

A major concern in the Hubbard system programming is the quantum particle type: we might know each site is $t(2)$ typed, only if we are given the physics background that the particles are fermions. Many physical systems, including the Hubbard model, are designed to be like an abstract class or a polymorphic higher-order function to be instantiated and used in dealing with different particles, e.g., bosons and qubits (see Section \ref{sec:boson}).
In \qsnd, we capture this polymorphism by typing on particle sites, where each particle site is classified as a specific $t(m)$ type, and different particle sites can have different types in \Cref{sec:typing}, so that every instance of a physical system has a type context, e.g., the Hubbard system above can be classified as a $t(2)$ typed function, i.e., $\sdag{a}(j):t(2) \to t(2)$ (we use a type construct $\quan{F}{\zeta}{t(2)}$ to represent the function type, see \Cref{sec:formal}).
This is particularly useful for transforming one particle system into another, often needed in analyzing quantum particle behaviors.
The compilation (\Cref{sec:boson}) of the $m$-particle bosonic Hubbard system to quantum computers is essentially a function having the following type, as defined in \qsnd.

{\small
\begin{center}
$
(\Motimes^m t(2^{n\splus 1}) \to \Motimes^{m} t(2^{n\splus 1})) \to (\Motimes^{m*(n\splus 1)} t(2) \to \Motimes^{m*(n\splus 1)} t(2))
$
\end{center}
}

%% file: semantics.tex
\section{Formalism}\label{sec:formal}

We discuss the semantics and type system of \qsnd by utilizing the Hubbard (\Cref{eq:hubbard}) model (system) as running examples.
To fully express a multiple particle state in Dirac notation, we can think of ket states for single particles are composed through tensor products as in \Cref{fig:data}, and superposition multiple particle state is expressed as a linear sum of different such ket tensor products.

\subsection{Semantics and Equivalence Rewrites}\label{sec:sem}

\begin{figure*}[t]
{\small
\begin{flushleft}\textcolor{blue}{Single Ket Semantics:}\end{flushleft}
{
\begin{center}
$
\denote{\sdag{a}}\ket{k}\rightarrow \sqrt{k\splus 1}\ket{k\splus 1}\;\;\cn{if}\;\;k\neq m
\qquad\quad
\denote{\sdag{a}}\ket{m}\rightarrow \zero
\qquad\quad
\denote{a}\ket{k}\rightarrow \sqrt{k}\ket{k\sminus 1}\;\;\cn{if}\;\;k\neq 0
\qquad\quad
\denote{a}\ket{0}\rightarrow \zero
$
\end{center}
}

  \ignore{
%\begin{flushleft}\textcolor{blue}{$\alpha$ Equivalence:}\end{flushleft}
%\begin{mathpar}
%    \inferrule[]{V(y,x,e)}
%                { \teq{\lambdae{x}{\omega}{e}}{\lambdae{y}{\omega}{e[y/x]}} }

%    \inferrule[]{V(f,g,e)}
%                { \teq{\smu{f}{\omega}{e}}{\smu{g}{\omega}{e[g/f]}} }
%  \end{mathpar}

\begin{flushleft}\textcolor{blue}{Type Equivalence:}\end{flushleft}
\begin{mathpar}
\hspace{-0.5em}
   \inferrule[]{}{ \teq{\sdag{\quan{F}{\zeta}{\iota}}}{\quan{F}{\zeta}{\iota}}}

   \inferrule[]{}{ \teq{t(\iota) \otimes t(\iota')}{t(\iota \otimes \iota')}}

   \inferrule[]{}{ \teq{\sdag{t}(\iota) \otimes \sdag{t}(\iota')}{\sdag{t}(\iota \otimes \iota')}}

   \inferrule[]{}{ \teq{\quan{F}{\zeta}{\iota} \otimes \quan{F}{\zeta}{\iota'}}{\quan{F}{\zeta}{\iota\otimes \iota'}}}
   
        \inferrule[]{}
        { \teq{\eexp{e}}{\cn{lfp}^{\infty}(n=0)(\frac{(\sapp{\sminus i}{e})^n}{n!})}}

     \inferrule[]{}
        { \teq{\elog{e}}{\sapp{-i}{\cn{lfp}^{\infty}(n=1)(\frac{(\sminus 1)(I - e)^n}{n}})}}

    \inferrule[]{}{ \teq{\sapp{e}{\sdag{(\lambdae{x}{\omega}{z * x})}}}{\sapp{\sdag{z}}{e}} }

    \inferrule[]{}
        { \teq{\cn{nor}(\sdag{e})}{\sdag{\cn{nor}(e)}} }
        
    \inferrule[S-Move]{}
        { \sapp{z a^{[\dag]}:\quan{F}{\zeta}{\iota}}{(z' \ket{j})} 
               \equiv (z*z') \sapp{a^{[\dag]}}{\ket{j}} }
  \end{mathpar}
}
  
\begin{flushleft}\textcolor{blue}{Semantics Rules:}\end{flushleft}
       \begin{mathpar}         
        \inferrule[S-Move]{}
        { (z a^{[\dag]}:\quan{F}{\zeta}{\iota},z' \ket{j}) \Downarrow
                (z*z') \denote{a^{[\dag]}}{\ket{j}}}
        
        \inferrule[S-Sum]{(e_1,\varphi)\Downarrow \varphi_1 \\ (e_2,\varphi)\Downarrow \varphi_2 }
        { ({e_1}\textcolor{spec}{+} {e_2},\varphi ) \Downarrow \varphi_1 + \varphi_2}
        
         \inferrule[S-Par]{(e,\varphi_1)\Downarrow \varphi'_1 \\ (e,\varphi_2)\Downarrow \varphi'_2 }
        { (e,\varphi_1 + \varphi_2 ) \Downarrow \varphi'_1 + \varphi'_2}

        \inferrule[S-App]{(e_2,\varphi)\Downarrow \varphi' \\ (e_1,\varphi')\Downarrow \varphi''}
        { (\sapp{e_1}{e_2},\varphi) \Downarrow \varphi''}
        
        \inferrule[S-Ten]{(e_1,w)\Downarrow \varphi \\ (e_2,w')\Downarrow \varphi'}
        { ({e_1}\textcolor{spec}{\otimes}\,{e_2},w\otimes w') \Downarrow \varphi \otimes \varphi'}
  \end{mathpar}
}
  \vspace*{-0.8em}
\caption{\qsnd Semantics. $a^{[\dag]}$ means either a creator $\sdag{a}$ or a annihilator $a$. The Marked \textcolor{spec}{blue} operations ($\textcolor{spec}{\otimes}$ and $\textcolor{spec}{+}$) are the ones in \qsnd, while marked black items are quantum state vector operations. } %$e^n = \underbrace{\sapp{e}{\sapp{e}{\sapp{\cdots}}{e}}}}_{n}$.}
\label{fig:equiv}
  \vspace*{-1.2em}
\end{figure*}

The \qsnd semantics is defined as an operational semantics in \Cref{fig:equiv}, with the judgment $(e, \varphi) \Downarrow \varphi'$.
Here, $e$ is the Hamiltonian program; $\varphi$ and $\varphi'$ are the pre- and post-quantum states.
The semantics captures the matrix meaning of a Hamiltonian, representing the snapshot effect of applying a boson-like particle system (\Cref{appx:twoviews}),
e.g., we apply a \qsnd Hamiltonian $e$ onto a quantum state $\varphi$.

The semantic function $\denote{-}$ in  \Cref{fig:equiv} captures the behaviors of applying bosonic creators and annihilators to a single ket of a single particle.
Here, $m$ is the upper bound of the basis vector for the ket state. When applying a creator, if the current basis-vector $k$ is not the upper bound, we increment the basis-vector and multiply an amplitude $\sqrt{k+1}$ to the vector. Essentially, the amplitude is pushed to the amplitude state in the ket containing the single basis vector, i.e., $z_1\eta_1 \otimes z_2\eta_2\equiv (z_1 * z_2)\eta_1 \eta_2 $ in \Cref{fig:data}.
If the basis vector is the upper-bound value $m$, we rewrite the basis vector to a $\zero$ vector. As we mentioned in \Cref{sec:overview}, we have $\zero \otimes \eta = \zero$, meaning that if a basis vector in a ket becomes $\zero$, the whole ket results in $\zero$.
Therefore, the $\zero$ vector behavior for creators and annihilators essentially eliminates the ket in the whole quantum state.
The annihilator's behavior is a mirror of the behavior of the creators.

Below the semantic function for creators and annihilators in \Cref{fig:equiv}, we provide the operational semantics for operations in \qsnd.
\rulelab{S-Move} applies creators (or annihilators) to the right basis-vector.
We can define the behavior of a single creator because we utilize our equational rules in \Cref{sec:typing} to push the $\dag$ operation to the single particle operation level.
\rulelab{S-Sum} defines the sum operation in \qsnd, i.e., we apply $e_1$ and $e_2$ both to the quantum state $\varphi$ and then linearly sum the result states together.
\rulelab{S-Par} defines the operation behavior for a \qsnd program $e$, if the quantum state is a superposition state.
Here, we apply $e$ to the two sides ($\varphi_1$ and $\varphi_2$) of the sum operation in the superposition state and then sum them together.

\rulelab{S-App} defines the function application as a matrix multiplication, and \rulelab{S-Ten} defines the tensor operation behavior on a ket state,
where we partition, via the tensor operation, the particle system into two parts and apply $e_1$ and $e_2$ to the two parts, respectively.
Part of the reason that we can describe the tensor semantics on the ket level is that we can use rule \rulelab{S-Par} to push a \qsnd program $e$ to each ket term in a superposition quantum state.
Another reason is that our type system ensures we can always make the correct partition, detailed in \Cref{sec:typing}.

With the semantic definition for the Hamiltonian, we can define the quantum simulation as follows.
Notice that the computation of the matrix exponent can be approximated as a Taylor series, which eventually produces another \qsnd program.
There are better compilation techniques to compile the Hamiltonian simulation to a quantum computer effectively, see \Cref{sec:compilation}.

\begin{definition}[Hamiltonian Simulation]\label{def:hamsim}\rm 
The simulation on a Hamiltonian $e$ with respect to time $t$ is defined as $\eexp{\sapp{e}{t}}$.
\end{definition}

%\liyi{provide an example.}

%\liyi{Discuss semantics and Hubbard model again by showing the fixed point and lambda definitions of the model.}

\subsection{Type System}\label{sec:typing}

The \qsnd typing judgment $\vdash e \triangleright \tau$ states that $e$ is well-typed as the type $\tau$, given in \Cref{fig:exp-proofsystem-1}.
The \qsnd equational rules examine the algebraic properties of \qsnd operations.
For simplifying the type system definition, we have a two-point lattice subtyping relation $\hmx\sqsubseteq \pmx$ among matrix operations.
The type system also enforces the three properties below.

\begin{figure*}[t]
  \vspace*{-0.5em}
{\footnotesize

\begin{flushleft}\textcolor{blue}{Expression Equivalence:}\end{flushleft}
\begin{mathpar}
    \inferrule[]{}
                { \teq{\sapp{(e_1 + e_2)}{e}}{(\sapp{e_1}{e}) + (\sapp{e_2}{e})} }

    \inferrule[]{}
               { \teq{\sapp{e}{(e_1+e_2)}}{(\sapp{e}{e_1}) + (\sapp{e}{e_2})} }

    \inferrule[]{}
                { \teq{{(e_1 + e_2)}\otimes{e}}{{e_1}\otimes{e} + {e_2}\otimes {e}} }
                        
    \inferrule[]{}
               { \teq{{e}\otimes{(e_1+e_2)}}{{e}\otimes{e_1} + {e}\otimes{e_2}} }
    
    \inferrule[]{}
        { \teq{\sdag{(e_1 \otimes e_2)}}{\sdag{e_1} \otimes \sdag{e_2}} }

    \inferrule[]{}
        { \teq{\sdag{(e_1 + e_2)}}{\sdag{e_1} + \sdag{e_2}} }
        
    \inferrule[]{}
        { \teq{ \sdag{(\sapp{e_1}{e_2})}}{\sapp{\sdag{e_2}}{\sdag{e_1}}}}          

    \inferrule[]{}
        { \teq{ \sdag{(\sdag{e})}}{e}}               
  \end{mathpar}

\ignore{
\begin{mathpar}
\hspace*{-1em}
    \inferrule[]{}
                { (\tau \otimes \tau_1) \to (\tau \otimes \tau_2) \equiv (\tau \to \tau ) \otimes (\tau_1 \to \tau_2)}

    \inferrule[]{}
                {\tau \otimes (\tau_1 \otimes \tau_2) \equiv (\tau \otimes \tau_1) \otimes \tau_2}

     % \inferrule[T-Var]{}
     %   {\tjudge{\Gamma}{x}{\Gamma(x)}}
        
    %\inferrule[T-Vec]{\forall j\in[0,n)\,.\,m_j < k}
    % {\tjudge{\Gamma}{\duala{j=0}{n}{m_j}{j}:t(n,k)}{(t(n,k))}}

  %  \inferrule[T-Lambda]{\tjudge{\Gamma[x\mapsto \omega]}{e}{\omega'}}
    %            {\tjudge{\Gamma}{\lambdae{x}{\omega}{e}}{\omega \to \omega'}}

  % \inferrule[T-Mu]{\tjudge{\Gamma[f\mapsto \omega \to \omega]}{e}{\omega}}
   %             {\tjudge{\Gamma}{\smu{f}{\omega}{e}}{\omega\to\omega}}
      % \inferrule[T-Dag]{\tjudge{\Gamma}{e}{t(\iota)}}
   %             {\tjudge{\Gamma}{\sdag{e}}{\sdag{t}(\iota)}}
   
      % \inferrule[T-Nor]{\tjudge{\Gamma}{e}{{t}^{[\dag]}(\iota)}}
    %            {\tjudge{\Gamma}{\cn{nor}(e)}{{t}^{[\dag]}(\iota)}}
                
    %\inferrule[T-Seq]{\tjudge{\Gamma}{e}{\quan{F}{\zeta}{\iota}}\quad\;\;\tjudge{\Gamma}{e'}{\quan{F}{\zeta'}{\iota}}}
    %            {\tjudge{\Gamma}{\sapp{e}{e'}}{\quan{F}{\zeta\sqcup \zeta'}{\iota}}}
                                
    %\inferrule[T-Mat]{\tjudge{\Gamma}{e}{\quan{F}{\zeta}{\iota}}\quad\;\;\tjudge{\Gamma}{e'}{\iota}}
    %            {\tjudge{\Gamma}{\sapp{e}{e'}}{\iota}}
                                
    %\inferrule[T-Inner]{\tjudge{\Gamma}{e}{\sdag{t}(\iota)}\\\tjudge{\Gamma}{e'}{t(\iota)}}
    %            {\tjudge{\Gamma}{\sapp{e}{e'}}{\mathbb{C}}}

   % \inferrule[T-Exp]{\tjudge{\Gamma}{e}{\quan{F}{\hmx}{\iota}}}
   %            {\tjudge{\Gamma}{\eexp{e}}{\quan{F}{\umx}{\iota}}}
                
    % \inferrule[T-Log]{\tjudge{\Gamma}{e}{\quan{F}{\umx}{\iota}}}
     %           {\tjudge{\Gamma}{\elog{e}}{\quan{F}{\hmx}{\iota}}}
  \end{mathpar}
  }
  
  \begin{flushleft}\textcolor{blue}{Typing Rules:}\end{flushleft}
  \begin{mathpar}
      \inferrule[T-Par]{e \equiv e' \\ \vdash e' \triangleright \tau}{ \vdash e \triangleright \tau }
    
    \inferrule[T-Op]{}
                {\tjudge{}{a:{\quan{F}{\pmx}{t(m)}}}{{\quan{F}{\pmx}{t(m)}}}}

    \inferrule[T-Dag]{\tjudge{}{e}{\quan{F}{\zeta}{\iota}}}
                {\tjudge{}{\sdag{e}}{\quan{F}{\zeta}{\iota}}}
                
    \inferrule[T-Tensor]{\tjudge{}{e}{\quan{F}{\zeta}{\iota}}\\\tjudge{}{e'}{\quan{F}{\zeta}{\iota'}}}
                {\tjudge{}{e \otimes e'}{\quan{F}{\zeta}{\iota \otimes \iota'}}}

    \inferrule[T-Plus]{\tjudge{}{e}{\tau}\\\tjudge{}{e'}{\tau}}
                {\tjudge{}{e + e'}{\tau}}
               
    \inferrule[T-App]{\tjudge{}{e}{\quan{F}{\zeta}{\iota}}\\\tjudge{}{e'}{\quan{F}{\zeta'}{\iota}}}
                {\tjudge{}{\sapp{e}{e'}}{\quan{F}{\zeta \sqcup \zeta'}{\iota}}}
                
    \inferrule[T-Her]{\tjudge{}{e}{\quan{F}{\pmx}{\iota}}\\\teq{\sdag{e}}{e} }
                {\tjudge{}{e}{\quan{F}{\hmx}{\iota}}}

  \end{mathpar}
}
  \vspace*{-0.8em}
\caption{Type rules. $\equiv$ is the equivalence relation. $\hmx\sqsubseteq \pmx$ and $\hmx\sqcup \pmx=\pmx$. }
\label{fig:exp-proofsystem-1}
  \vspace*{-1.2em}
\end{figure*}

%Guaranteeing Program Wellformedness and 
\myparagraph{Typing Lattice Sites.}
In a second quantization lattice graph structure, each particle site might have different state types ($t(m)$), and each site state is a $m$ dimensional Hilbert space. The correctness of our type system ensures proper typing.
In applying to a single particle site, rule \rulelab{T-Op} ensures that a singleton creator or annihilator refers to the proper state type ($t(m)$).
We also ensure that the function flag is $\pmx$ because a singleton operation is neither Hermitian nor unitary.

Each particle site might have a different type, connected by a tensor operation ($\otimes$), typed by rule \rulelab{T-Tensor}.
For example, One can construct a two particle system $v_1:t(m) \otimes v_2:t(j)$ and applying two annihilators $a \otimes a$ to the system,
where the two annihilators ($a$) are typed as $\quan{F}{\zeta}{t(m)}$ and $\quan{F}{\zeta}{t(j)}$, respectively.
The typing clarifies different particle types so one can refer to the types for analyses.
The result type of the tensor becomes $\quan{F}{\zeta}{t(m) \otimes t(j)}$.
We also ensure the correct type in a quantum choice operation, i.e., rule \rulelab{T-Plus} implies that quantum superposition is a quantum version of choice operators. The two sides, $e$ and $e'$, are required to output the same type.

\myparagraph{Type Switching of Different Matrix Operations.}
To ensure that we can use \qsnd to define a Hamiltonian (Hermitian matrix) correctly, we develop type flags $\pmx$ and $\hmx$, representing ordinary and Hermitian (Hamiltonian), respectively.
The subtyping relation $\hmx\sqsubseteq \pmx$ subtypes an $n$-dimensional Hamiltonian or unitary as an $n$-dimensional ordinary matrix.
Rule \rulelab{T-Her} lifts an ordinary matrix to a Hamiltonian by checking if $e$ and $\sdag{e}$ are equivalent via our equational rewrite rules.
For example, in the Hubbard system in \Cref{sec:overview}, the term $\sapp{\sdag{a}(j)}{ a(j\splus 1)} + \sapp{\sdag{a}(j\splus 1)}{ a(j)}$ defines a Hamiltonian because its conjugate transpose is equivalent to itself. 

\myparagraph{Equational Properties.}
Equational rules in \Cref{fig:exp-proofsystem-1} examine the algebraic properties in a \qsnd program.
They help push the $\dag$ operations around the single particle operations so we can define the behavior of $\dag$ by only specifying the semantics of creators.
Additionally, we can use the $\dag$ rule to check if a \qsnd program is Hermitian, which is essential to generate proper quantum unitary programs in Hamiltonian simulation, i.e., in \Cref{def:hamsim}, the exponent produces a unitary, if only if $e$ is Hermitian.

\ignore{
\myparagraph{Permitting Equivalence Relations.}
The type system enables the equivalence relations, shown in \Cref{fig:equiv}, that support the semantic definitions in \Cref{fig:sem}.
The first line performs $\alpha$ conversions in a $\lambda$ and a $\mu$ term, provided that the substituted variable does not cause variable capturing issues.

    %\inferrule[]{}
    %            { \teq{\sapp{(e_1 \otimes e_3)}{(e_2 \otimes e_4)}}{(\sapp{e_1}{e_2}) \otimes (\sapp{e_3}{e_4})} }

    %\inferrule[]{} maybe big mistakes to include the rules that are commented
    %            { \teq{}{\lambdae{x}{\tau_1 \otimes \tau_2 \to \tau_1 \otimes \tau_2}{e_1 \otimes e_2}}{ 
    %             (\lambdae{x}{\tau_1\to \tau_1}{e_1}) \otimes (\lambdae{x}{\tau_2\to \tau_2}{e_2})} }
    %\inferrule[]{}
    %            { \smu{f}{\tau_1 \otimes \tau_2}{e_1 \otimes e_2} \equiv 
    %             (\smu{f}{\tau_1}{e_1}) \otimes (\smu{x}{\tau_2}{e_2}) }
        %\inferrule[]{}
    %            { \lambdae{x}{\tau}{e_1 + e_2} \equiv (\lambdae{x}{\tau}{e_1}) + (\lambdae{x}{\tau}{e_2})}

    %\inferrule[]{}
    %            { \smu{x}{\tau}{e_1 + e_2} \equiv (\smu{x}{\tau}{e_1}) + (\smu{x}{\tau}{e_2})}
    
The second line relates application operations with different particle sites, separated by tensor operations ($\otimes$).
The equivalence relation implements distributed laws. The left-hand rule deals with the case when $e_1$ and $e_3$ are row vectors, and the right-hand rule happens when $e_1$ and $e_3$ are matrix operations ($\quan{F}{\zeta}{\kappa(n)}$).
The result pairs up $e_1$ and $e_2$ as well as $e_3$ and $e_4$, with the connection of a tensor operation.
In the two rules, the left-hand side can be rewritten to the right-hand side if $e_1$ and $e_2$ are typed to have the same dimensionality.

The third line deals with sum operations. When dealing with a sum of kets, we can move the operation $e$ to apply to the kets directly.
The right-hand-side rule distributes the superposition sum operations $e_1$ and $e_2$, so the rewrite results in a sum of two applications of applying $e_1$ and $e_2$ to the state $e_3$.
The fourth line deals with transpose operations ($\dag$), where the operation moves to the inner level when it is combined with a tensor ($\otimes$) and sum ($+$) operation.
When combined with an application, the order of the application is swapped if $e_1$ is typed as a row vector or a matrix operation, while the order is fixed if $e_1$ is typed as a number.
The last line defines the semantics for matrix exponential and logarithm operations as equivalence relations.
They are defined as a power series, which may converge to a fixed matrix operation. The exponential of a Hamiltonian should converge to a unitary, while the logarithm of a unitary should converge to a Hamiltonian. The \qsnd compilation implements the two operations as finite approximations.

\liyi{Maybe add some examples. }
}

\myparagraph{Metatheories.}\label{sec:theorem}
We show the \qsnd type soundness theorem, stating that a given \qsnd program applies to a well-typed quantum state does not get stuck, and the resulting state is also well-typed.
A well-typed state $\iota \vdash \varphi$ means that each particle site described by $\varphi$ respects the particle site types indicated in $\iota$.
%
%The result relies on the definition of a \emph{value} ($\nu$), which extends the value definition in $\lambda$ calculus (a variable $x$ or a lambda/mu abstraction) to include pure quantum states $\psi$, row vectors $\sdag{\psi}$, numbers subtyped to $\mathbb{C}$, and pure matrix operations ($\quan{F}{\xi}{\iota}$) consisting only of creators, annihilators, linear sums, and tensors.

%\vspace*{-0.2em}
\begin{theorem}[Type Soundness]\label{thm:type-progress-oqasm}\rm 
  If $\vdash e \triangleright \quan{F}{\xi}{\iota}$, and $\iota \vdash \varphi$, then $(e,\varphi) \Downarrow \varphi'$ and $\iota \vdash \varphi'$.
\end{theorem}

Based on \Cref{def:hamsim}, the Hamiltonian simulation, $\eexp{{\hat{H}}{t}}$, of a Hermitian Hamiltonian $\hat{H}$ produces a unitary.
Our type soundness theorem in \Cref{thm:type-progress-oqasm} shows that every Hermitian type ($\hmx$) program $e$ can be turned into a unitary through the matrix exponent operation ($\eexp{e}$).
Additionally, Deutsch \textit{et al.} \cite{Deutsch1989QuantumCN,Deutsch1995} showed that any unitary matrix can be decomposed to elementary quantum gates, executable in quantum computers.
Combined with our type soundness theorem, this fact results in the following compilation existence corollary.

\begin{corollary}[Compilation Existence]\label{thm:compile-good}\rm 
There is a compiler to compile any $\hmx$ flagged \qsnd program to digital quantum computers.
\end{corollary}

The theorem does not indicate an efficient compilation. We discuss such compilation in \Cref{sec:compilation}.

%% file: compilation.tex
\section{Compilation}\label{sec:compilation}

Here, we discuss the compilation of the Hamiltonian simulations of \qsnd programs to quantum computers. 
%A discussion about the compilation for performing expectation value calculation is in \Cref{sec:groundenergy}.
We then show a case study on compiling Boston particle systems.
We implement a Python compiler to compile a Hamiltonian simulation and energy state computation via the mechanisms mentioned in \Cref{sec:qcompile}, and we use our Python compiler to execute all the examples and case studies.

\subsection{Compilation to Quantum Computational Platforms}\label{sec:qcompile}

\begin{figure}[t]
\vspace*{-0.5em}
{\begin{center}
  \includegraphics[width=0.7\textwidth]{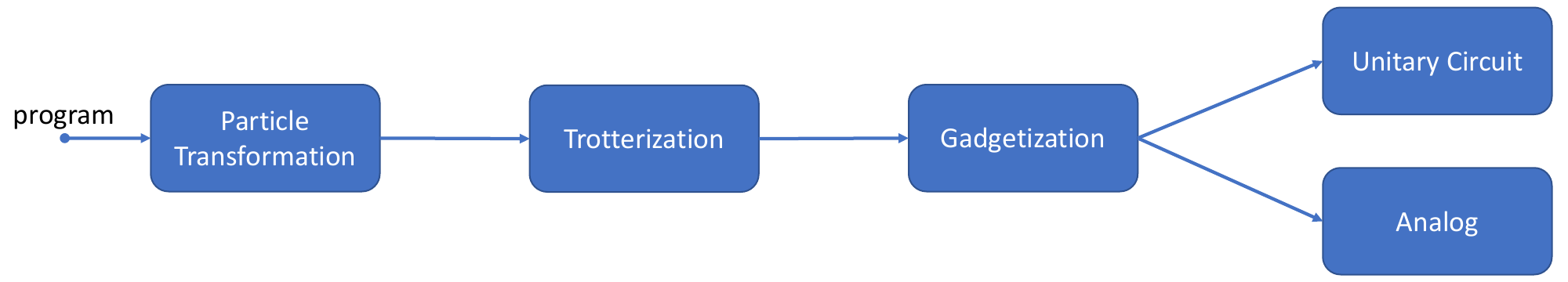}
  \end{center}
 }
  \vspace*{-0.5em}
     \caption{Quantum Compilation Flow}\label{fig:fig-compilation}
\vspace*{-0.5em}
\end{figure}

The execution of a \qsnd program typically depends on the specific applications, e.g., time-propagated Hamiltonian simulation and energy state calculation, mentioned in \Cref{sec:background}. Here, we discuss the effective compilation of Hamiltonian simulation.

There are several well-known compilation strategies \cite{Low_2019,Low_2017,Berry_2006,10.5555/2481569.2481570,Berry_2015,Berry_2015a}, summarized as the two compilation procedures, digital- and analog-based compilations, in \Cref{fig:process}.
The compilation flow of a Hamiltonian simulation to digital and analog-based quantum computers is given in \Cref{fig:fig-compilation}.
Hamiltonian simulation and digital-based quantum computers are essentially two viewpoints of the same concept, and digital-based quantum computers are implemented in quantum simulators \cite{Tacchino_2019,Li_2022}. 
The two procedures have the same first two transformation steps, second quantization transformation and Trotterization \cite{Lloyd96}, and have some differences in the final circuit generation steps. We examine all of them below.

\noindent{\textbf{{Second Quantization Transformations.}}}
Quantum computers are essentially a $t(2)$ typed Hamiltonian particle system.
There are two observations for the current quantum computers: 1) they are lattice-based particle systems with particle sites having type $t(2)$, i.e., a singleton particle site is a qubit; and 2) the simulation result must be unitary in a quantum computer, meaning that the input program must be Hermitian; thus, creators and annihilators cannot be directly implemented since they individually are not Hermitian. Quantum computers usually permit the definitions of Pauli groups below, individually being Hermitian, as an alternative representation of $t(2)$ typed creators and annihilators.

{\small
\[
I \triangleq  
\begin{bmatrix}
1 & 0\\
0 & 1
\end{bmatrix}
\qquad
X \triangleq  
\begin{bmatrix}
0 & 1\\
1 & 0
\end{bmatrix}
\qquad
Y \triangleq  
\begin{bmatrix}
0 & \sminus i\\
i & 0
\end{bmatrix}
\qquad
Z \triangleq  
\begin{bmatrix}
1 & 0\\
0 & \sminus 1
\end{bmatrix}
\]
}

The meaning of Pauli groups in terms of second quantization differs from the gates in a digital quantum circuit, which provides a different view of particle movements in a Hamiltonian. Instead of describing electron occupation by creators and annihilators, Pauli groups describe particle movements as rotations across \(X\), \(Y\), and \(Z\) bases.
There are different transformation methods. Here, we define the Pauli group operations as the $t(2)$ typed creators and annihilators below; such definition is based on the bosonic Holstein–Primakoff transformation \cite{qua.25176} (or Jordan-Wigner Transformation for fermions, see \Cref{sec:jw-trans}).

{
\[
I = \sapp{\sdag{a}}{a} + \sapp{a}{\sdag{a}} 
\qquad
X = \sdag{a} + a
\qquad
Y = i a - i \sdag{a}
\qquad
Z = \sapp{\sdag{a}}{a} - \sapp{a}{\sdag{a}} 
\tag{3.3.1}\label{eq:pauli}
\]
}

Here, \(i\) is a complex amplitude and \(i a - i \sdag{a}\) is an abbreviation of \(\sapp{\sdag{a}}{a} + (\sapp{(\sminus 1)a}{\sdag{a}})\). 
To understand the transformation, recall that the matrix representation of creators \(\sdag{a}\) and annihilators \(a\) are $\begin{psmallmatrix}
0 & 0\\
1 & 0
\end{psmallmatrix}$ and $\begin{psmallmatrix}
0 & 1\\
0 & 0
\end{psmallmatrix}$, respectively, and a Hamiltonian represents a snapshot showing the potential movement of a particle. 
Defining the \(X\), \(Y\), and \(Z\) bases in a second quantization formalism identifies three different orthonormal bases, representing different ways of potential movements. For example, the \(X\) basis is a sum of a creator and annihilator. If we view a sum operation as a quantum choice, the \(X\) basis transformation indicates that a potential movement in an \(X\) basis is either to create or annihilate an electron. The $Y$ and $Z$ bases are defined to ensure the orthonormality between them and against the \(X\) axis. 

The above transformation only transforms a \qsnd Hamiltonian describing only $t(2)$ typed particles to a program with only Pauli operations.
The transformation is much more complicated than other kinds, e.g., $t(m)$ typed particles. We show an example in \Cref{sec:boson}.

\noindent{\textbf{{Trotterization.}}}
The second quantization transformation above produces a \qsnd program in the form of Pauli operations.
To compile a \qsnd program with many linear sum terms, we perform Trotterization, based on the Lie-Trotter product formula (left below) \cite{Lie1880} and the approximated formula, on the right, developed by Suzuki \cite{529425}.

{\small
\begin{center}
$\cn{exp}(A+B)=\lim_{n\to \infty}\left(\sapp{\cn{exp}({\frac{A}{n}})}{\cn{exp}({\frac{B}{n}})}\right)^n
\qquad
\qquad
\cn{exp}(\delta(A+B))=\sapp{\cn{exp}(\delta(A))}{\cn{exp}(\delta(B))}+O(\delta^2)
$
\end{center}
}

For a time period $t$ and Hamiltonian $\hat{H}$, Trotterization finds a set of small Hamiltonians $\hat{H}=\sum^n_{j=1} \hat{H}_j$, each of which is effectively simulatable in quantum computers, thus, $\eexp{\hat{H}_1 t}...\eexp{\hat{H}_n t}$ approximates $\eexp{\hat{H} t}$. Each matrix exponential operation generates a unitary, and Trotterization sequentializes these generated small unitaries.
The Suzuki approximation above suggests that the error rate $O(\delta^2)$ is related to the square of the variable term $\delta$.
In the context of Hamiltonian simulations, $\delta$ refers to the $\sminus \sapp{i}{t}$, suggesting that the error rate is related to the selected period.
Thus, to simulate a quantum system with respect to a period $t$, one can select a large $n$ to consecutively apply the exponent $\sapp{\cn{exp}({\frac{A}{n}})}{\cn{exp}({\frac{B}{n}})}$ to the state $n$ times, so $\frac{t}{n}$ diminishes such that the error rate $O({(\frac{t}{n})}^2)$ is negligible.
Such tactics can be generalized to arbitrary numbers of summation terms.

{\small
\[
\hat{H}_{I1} = \sum_{j} \sapp{Z(j)}{Z(j\splus 1)} + z_h \sum_j X(j)
\]
}

\begin{figure*}[t]
{\small
\begin{tabular}{c@{$\;\;=\;\;$}c}
\begin{minipage}{0.22\textwidth}
$
\eexp{\sapp{\sapp{Z(j)}{Z(j\splus 1)}}{\frac{t}{n}}}
$
\end{minipage}
&
\begin{minipage}{0.3\textwidth}
{
  \Qcircuit @C=0.5em @R=0.5em {
    \lstick{} & & \ctrl{1}       & \qw  & \ctrl{1}      &\qw\\
    \lstick{} & & \targ         & \gate{Rz(2\frac{t}{n})}       & \targ    & \qw 
    }
}   
\end{minipage}
\end{tabular}
\hspace{1em}
\begin{minipage}{0.41\textwidth}
{\small
$
ZZ(\theta)=
\begin{psmallmatrix}
\cn{exp}(\sminus i\frac{\theta}{2}) & 0 & 0 & 0\\
 0 & \cn{exp}(i\frac{\theta}{2}) & 0 & 0\\
 0 & 0 & \cn{exp}(i\frac{\theta}{2}) & 0\\
 0 & 0 & 0 & \cn{exp}(\sminus i\frac{\theta}{2})
\end{psmallmatrix}
$
}
  \end{minipage}
}
\vspace*{-0.5em}
\caption{Digital circuit gates and unitary for $ZZ$ Interaction. The left applies on $j$ and $j\splus 1$ qubits.}
\label{fig:zz-inter}
\vspace*{-0.3em}
\end{figure*}

For example, to simulate $\hat{H}_{I1}$ \footnote{The $\hat{H}_{I1}$ equation is a special $1D$ Ising model with uniformed interaction strength, where the first terms in the quantum choice represent the particle interactions. In contrast, the second terms represent external transverse magnetic field effects, with $z_h$ being the external transverse magnetic field strength.} above with a time slot $t$ through the pick of a splitting number $n$, we compute the exponent $\eexp{\sapp{\hat{H}_{I1}}{t}}$ using the approximation as follows.

{\small
\begin{center}
$\eexp{\sapp{\hat{H}_{I1}}{t}}
= \left(\eexp{\sapp{(\sum_{j} \sapp{Z(j)}{Z(j\splus 1)} + z_h \sum_j X(j))}{\frac{t}{n}}}\right)^n
= \left(\sapp{\sapp{\eexp{\sapp{\sapp{Z(1)}{Z(2)}}{\frac{t}{n}}}}{...}}{\sapp{\eexp{\sapp{X(1)}{\frac{t}{n}}}}{...}} \right)^n
$
\end{center}
}

For example, one small sum term in the Hamiltonian is $\sapp{Z(j)}{Z(j\splus 1)}$, applying on $j$-th and $j\splus 1$-th qubits, with respect to a tiny period $\frac{t}{n}$. Trotterization tries to find a large number $n$ so the simulation of many small periods $\frac{t}{n}$ can correctly approximate the simulation of the original system Hamiltonian.
The result of gate synthesizing on the machine Hamiltonian, as to find gates representing $\eexp{\sapp{\sapp{Z(j)}{Z(j\splus 1)}}{\frac{t}{n}}}$, is shown as the left-hand side in \Cref{fig:zz-inter}, performing the digital circuit program, performing two controlled-not gates and an $\cn{Rz}(2\frac{t}{n})$ gate in the middle for the $j$-th and $j\splus 1$-th qubits,
i.e., $\cn{Rz}(\theta)$ is an $z$-axis phase rotation gate, rotating $\theta$ angle.

\begin{figure}
{\small
\begin{center}
$
\begin{array}{c}
{\eexp{\sapp{\theta}{X}}}=\cn{RX}(2{\theta})
=
\begin{psmallmatrix}
\cos{{\theta}} & -i\sin{{\theta}}\\
-i\sin{{\theta}} & \cos{{\theta}}
\end{psmallmatrix}
\qquad
{\eexp{\sapp{\theta}{Y}}}=\cn{RY}(2{\theta})
=
\begin{psmallmatrix}
\cos{{\theta}} & -\sin{{\theta}}\\
\sin{{\theta}} & \cos{{\theta}}
\end{psmallmatrix}
\\[0.3em]
{\eexp{\sapp{\theta}{Z}}}=\cn{RZ}(2{\theta})
=
\begin{psmallmatrix}
\eexp{{\theta}} & 0\\
0 & \cn{exp}{(i {\theta})}
\end{psmallmatrix}
\end{array}
$
\end{center}
}
\vspace*{-1em}
  \caption{Pauli matrices simulated to rotation gates. $\eexp{\sapp{\theta}{P}}$ means simulating $P$ on time $\theta$.}
  \label{fig:paulisim}
  \vspace*{-1em}
\end{figure}

\noindent{\textbf{{Gadgetization.}}}
After the Trotterization step, the digital-based compilation performs gate syntheses (Gadgetization), while analog-based compilation tries to cut the above small Hamiltonians even smaller to satisfy the target machine limitation. For gate syntheses, in the Bloch qubit sphere interpretation \cite{mike-and-ike}, the rotation gate $\cn{Rx}$, $\cn{Ry}$, $\cn{Rz}$ can be expressed as some exponents of Pauli matrices, which essentially suggests that the simulation of singleton Pauli $X$, $Y$, and $Z$ terms are the above phase rotation gates, shown in \Cref{fig:paulisim}.
For example, the gate synthesis of $\eexp{\sapp{X(j)}{\frac{t}{n}}}$ generates an $\cn{Rx}$ gate, as $\cn{Rx}(\frac{2t}{n})(j)$. The gate synthesis of the $z$-axis interaction exponent $\eexp{\sapp{\sapp{Z(j)}{Z(j\splus 1)}}{\frac{t}{n}}}$ is introduced in \Cref{fig:zz-inter}.
Ultimately, the simulation of the whole $\hat{H}_{I1}$ system is a series of applications of the two gates above.

The Gadgetization step has different compilations for digital- and analog-based machines.
We now show the difference in the step for analog-based compilation but can also appear in the digital-based compilation, i.e., the main goal of the pulse synthesis Gadgetization step in a digital-based compilation is to fit target machine Hamiltonians \cite{10.1145/3632923,Li_2022,Tacchino_2019}.
Essentially, quantum gates are simulated in the underlying quantum machines because they are essentially Hamiltonian simulators (\Cref{appx:gatesyn}).
Instead of performing quantum gate syntheses, one can generate good small Hamiltonians that can be directly implemented as specific target machine Hamiltonians and perform machine-level simulations.
For example, an IBM machine is based on the Heisenberg system \cite{Auerbach1998-jd} and permits the following machine Hamiltonian between any of the two adjacent sites $j$ and $j\splus 1$.

{\small
\[
\hat{H}_{\cn{IBM}} = z_1\sapp{Z(j)}{X(j\splus 1)} + z_2\sapp{Z(j)}{Z(j\splus 1)}+z_3 Z(j) + z_4 X(j\splus 1)
\]
}

Since users have the freedom to set the values $z_1$, $z_2$, $z_3$, and $z_4$, this indicates that an IBM machine permits $ZX$ and $ZZ$ interactions as well as $Z$ and $X$ axis rotations. 
To simulate a simple Ising system $\hat{H}_{I1}$, for each $j$-th position, we can set the values $z_1$ and $z_3$ to be $0$, $z_2$ to be $1$, and $z_4$ to be $z_h$.
This results in a more efficient compilation.
The above fitting step essentially implements the $ZZ$-interaction gate in \Cref{fig:zz-inter}, and also shows the reason for some interaction gates being much more efficient than elementary gates in IBM machines, evidenced in \cite{10.1145/3632923}.
Other quantum machines have different machine Hamiltonian representations, such as Quera machines \cite{quera2024}, and the machine Hamiltonian fitting step needs to perform different fitting methods for different quantum machines.
%The Ising system simulation is a toy example. \Cref{sec:boson} presents another example involving the simulation of a bosonic system.

%\subsection{Behind Quantum Circuits: Quantum Computers are Hamiltonian Simulators}\label{sec:sim-gates}

%% file: case.tex
%\vspace{-0.3em}
%\section{Case Studies}\label{sec:cases}

%In this section, we show two case studies for the utility of \qsnd.

\ignore{
\subsection{The Quantum Interpretation of Clique Finding}\label{sec:clique} 

\begin{figure}[t]
{\hspace*{-1.3em}
\begin{tabular}{c@{\quad}c}
\begin{minipage}[b]{0.81\textwidth}
  \includegraphics[width=\textwidth]{adiabatic}
  \vspace*{-1.3em}
  \caption{Adiabatic Energy State Evolution to Find a Clique}
\label{fig:adiabatic}
\end{minipage}
&
\begin{minipage}[b]{0.19\textwidth}
  \includegraphics[width=\textwidth]{case1.pdf}
  \vspace*{-1.3em}
  \caption{Clique Algo.}
\label{fig:case1}
\end{minipage}
\end{tabular}
}
\vspace*{-1.3em}
\end{figure}

We present a quantum algorithm to find the largest clique in a random graph based on the adiabatic evolution concept \cite{farhi2000quantum}, the quantum version of the constraint Hamiltonian energy programming in \Cref{sec:overview1}, which is similar to the Hamiltonian simulation in \Cref{sec:background}, however, with a different interpretation.
Here, the input state $\psi(0)$ is typically a superposition of states encapsulating all possible ground eigenstates, and we define a Hamiltonian to represent the objective function and embed the constraints we want to impose on the system.
The adiabatic evolution methodology uses the constrained Hamiltonian to derive the convergence of the quantum states towards the desired ground state.
Upon performing the Hamiltonian simulation by allowing the Hamiltonian to evolve for a considerably long time period $T$, the ground state is eventually reached.

As indicated in the analogy in \Cref{fig:adiabatic}, in computing the largest clique in a random graph through the adiabatic evolution, we first initialize a superposition state representing the approximate eigenstates for all possible cliques in a graph, analogous to the colored region in the figure.
We then implement a constraint by adding penalties for edges not connected in an input graph. For instance, the $+1$ edges in \Cref{fig:adiabatic}, representing the disconnected edges, will increase the total energy. At the end of the simulation, we ideally expect the system to resolve into a state that avoids such edges in the solution.
Each step of the adiabatic evolution effectively attempts to remove unsatisfied vertices so that the minimized group state, without penalties, can be reached; such a procedure is analogous to the transitions in \Cref{fig:adiabatic}.
The power of quantum adiabatic evolution is that the initial superposition state can simultaneously allow different tries, as the choices in the search list in the classical Hamiltonian energy computation of finding a $k$-clique. These tries happen simultaneously, and high-energy states (bad states) collapse to low-energy states in each transition, approaching the final ground state.

{
\begin{center}
$
\hat{H}(s) = (1-\frac{s}{T})\hat{H}_B + \frac{s}{T}\hat{H}_P
\qquad
\underbrace{\eexp{\hat{H}(0) t}\eexp{\hat{H}(t) t}...\eexp{\hat{H}(n*t) t}}_{n}
$
\end{center}
}

The algorithm procedures are given in \Cref{fig:case1}.
The whole algorithm is a loop structure. We need to select a large enough time period $T$ so the proper eigenstate $\psi(T)$, representing the largest clique, can be obtained.
We split the time period into small segments. For each segment $t\in[0,T]$, we simulate $\eexp{\sapp{(\hat{H}(t))}{t}}$, as the left equation above.
If the time segments are selected evenly, the number of Hamiltonian simulation iterations is $n=\frac{T}{t}$, given in the equation on the right.

{\footnotesize
\begin{center}
$
\hat{H}_B = -\sum_{j,k} \sapp{\sdag{a}(j)}{a(k)}+\sapp{a(j)}{\sdag{a}(k)}
\quad
\hat{H}_P = -\sum_{j,k} (1\sminus G(j,k))\sapp{\sapp{\sdag{a}(j)}{a(j)}}{\sapp{\sdag{a}(k)}{a(k)}}
\quad
\psi_0 = 
\begin{pmatrix} n \\ k \end{pmatrix}^{-\frac{1}{2}}\sum\ket{z_k}^{\otimes n}
$
\end{center}
}

The Hamiltonian, representing a graph in an adiabatic evolution, can be decomposed as two parts, $\hat{H}_B$ and $\hat{H}_P$, conceptually similar to the clique Hamiltonian at the end of \Cref{sec:overview1}.
The initial Hamiltonian $\hat{H}_B$ represents our initial guess of a clique in the input graph where all edges are connected, which is the lowest energy expectation value in the system. The constrained evolving Hamiltonian $\hat{H}_P$ represents a penalty adding to the system if an edge is missing between $j$-th and $k$-th vertices.
The concept of adding penalties is similar to adding repulsion penalties in the Hubbard system (\Cref{eq:hubbard}), e.g., we utilize the combination of creators and annihilators to remove ket states other than the ones with two $1$s in the $j$-th and $k$-th vertex locations, meaning that we want to add a penalty if a subgroup contains the $j$-th and $k$-th vertices and they are disconnected, constrained by $1-G_{j,k}$.
The initial state $\psi_0$ is a superposition state of all the $k$-subgraph vertex sets ($k \le n$), similar to the one below, where $m_k=1$ if the vertex $k$ is in the subgraph.
If we model a basis vector $\ket{m_0}\ket{m_1}...\ket{m_{n\sminus 1}}$; each $m_k$ represents a vertex in the graph.

{\small
\begin{center}
$\begin{pmatrix} n \\ k \end{pmatrix}^{-\frac{1}{2}}
(
\ket{\underbrace{0...0}_{n}}
+
\ket{1\underbrace{0...0}_{n\sminus 1}}
+
\ket{01\underbrace{0...0}_{n\sminus 2}}
+
...
+
\ket{11\underbrace{0...0}_{n\sminus 2}}
+
...
+
\ket{\underbrace{11....1}_{n}}
)
$
\end{center}
}
%\vspace{-0.5em}

There are $\begin{pmatrix} n \\ k \end{pmatrix}$ different combinations of such vertex sets to form subgraphs.
The state preparation of such a superposition state can be another Hamiltonian simulation process mentioned in the work of \cite{10.5555/2011430.2011431}.
Through simulating (evolving) the Hamiltonian long enough, the system reaches its lowest energy state, the eigenstate $\psi(T)$. Upon measurement, it results in a candidate maximal clique in the graph, if there is any.
Unfortunately, it is impossible to foreknow the value to be used for $T$ for adiabatic evolution; therefore, we have a loop structure in \Cref{fig:case1} to try different values for $T$.
Subsequent iterations increase $T$ progressively to see if the system converges to a valid ground state.
There are several ways of executing adiabatic simulations.
One such is by reformulating the Hamiltonian to the Ising model through quantum annealing machines \cite{Kadowaki_1998}.  
The circuit compilation of the exponent terms is another mechanism for performing adiabatic simulations. It is the same procedure as the one described in the next section.
}

\subsection{Compiling Boson Particle Systems to Quantum Computers}\label{sec:boson} 

We show the compilation of a bosonic particle system to quantum computers as a case study.
Bosons are a special kind of particle and typically require infinite-dimensional Hilbert space to describe a particle, e.g., a boson is typed $t(\infty)$.
A typical way of approximating a bosonic system \cite{qua.25176} utilizes $2^{n\splus 1}$ dimensional Hilbert space, i.e., they use a length $n$ basis vector to track the particle state, representing a possible spot (orbital) for a particle to occupy, e.g., $\sdag{a}\ket{j}=\ket{j\splus 1}$ means that we zero the $j$-th position and mark the $j\splus 1$-th bit to be $1$.
The extra $n$-th position acts as an overflow bit, indicating if a maximal orbital is reached.
Essentially, the transformation from an $m$-particle bosonic system to a qubit system --- the Holstein–Primakoff transformation for bosons \cite{qua.25176}) --- is a function $\gg^m:\quan{F}{\zeta}{\Motimes^m t(2^{n\splus 1})} \to \quan{F}{\zeta}{\Motimes^{m*(n\splus 1)} t(2)}$, compiling a $t(2^{n\splus 1})$ typed system to a $t(2)$ typed one, where the $t(2)$ typed system is written in the form of Pauli groups.
Since the bosonic system transformation is compositional, it is enough to show the transformation for one particle case $\gg:\quan{F}{\zeta}{t(2^{n\splus 1})} \to \quan{F}{\zeta}{t(2)}$, and view the $m$-particle case ($\gg^m$) as a map function by mapping $\gg$ to each particle in the system.
Apparently, the numbers of particles in the $t(2^{n\splus 1})$ and $t(2)$ typed systems ($m$ vs. $m*(n\splus 1)$) are different; thus, the particle site positions need to be rearranged.
In compiling a $t(2^{n\splus 1})$ system, we assume that a $j$-th position operation $op(j)$ is transformed to operations indexed $(j,0) ... (j,n)$, as $op'_0(j,0) ... op'_n(j,n)$, where the actual implementation of the $2D$ indices can be easily handled by a higher order function.
To describe the transformation function $\gg$, it is enough to show the transformation of a $\quan{F}{\pmx}{t(2^{n\splus 1})}$ typed creators and annihilators below.

{\small
\begin{center}
$
\begin{array}{l@{\;\;}c@{\;\;}l@{\;}c@{\;}l}
 \sdag{a}(k):\quan{F}{\pmx}{t(2^{n\splus 1})} &\gg& \sum_{j=0}^{n\sminus 1}\sqrt{j\splus 1}(\sdag{a}\textcolor{spec}{(k,j)}:\quan{F}{\pmx}{t(2)} &\otimes& {a}\textcolor{spec}{(k,j\splus 1)}:\quan{F}{\pmx}{t(2)})
\\[0.4em]
a(k):\quan{F}{\pmx}{t(2^{n\splus 1})} &\gg& \sum_{j=0}^{n\sminus 1}\sqrt{j\splus 1}(a\textcolor{spec}{(k,j)}:\quan{F}{\pmx}{t(2)} &\otimes& \sdag{a}\textcolor{spec}{(k,j\splus 1)}:\quan{F}{\pmx}{t(2)})
\end{array}
$
\end{center}
}

The indices in \textcolor{spec}{blue} above only serve indicator purposes.
The compilation of $t(2)$ typed creators and annihilators to Pauli operations is a direct result of \Cref{eq:pauli}.

{\small
\begin{center}
$
\sdag{a}:\quan{F}{\pmx}{t(2)}= \frac{1}{2}(X + i Y)
\qquad
a:\quan{F}{\pmx}{t(2)}= \frac{1}{2}(X - i Y)
$
\end{center}
}

We now see an application of transforming a two-particle bosonic system, with $n=1$, to quantum computers for simulating the boson behavior based on the Bose-Hubbard system \cite{peng2023quantum}, an instantiation of the Hubbard system (as an abstract class) \Cref{eq:hubbard} to describe the superconducting behaviors of boson particles.
Similar to the Hubbard system, a two-particle Bose-Hubbard system comprises of an expression $\sapp{\sdag{a}(0)}{a(1)}+\sapp{a(0)}{\sdag{a}(1)}$, referring to that index $0$-th and $1$-st particles might be exchanged in the next moment. The system also contains the repulsion term, similar to the one in \Cref{eq:hubbard}.

We now see how to compile a subterm of the two-particle system, $\sapp{\sdag{a}(0)}{a(1)}+\sapp{a(0)}{\sdag{a}(1)}$, to quantum computers.
Such an expression is syntactic sugar, discussed in \Cref{sec:runningexample}.
A simplified version of the expanded expression is ${\sdag{a}\textcolor{spec}{(0)}}\otimes {a\textcolor{spec}{(1)}}+{a\textcolor{spec}{(0)}}\otimes {\sdag{a}\textcolor{spec}{(1)}}$, obtained by distributing the tensor products to show the position relations among different operations,
i.e., $(\sapp{(\sdag{a} \otimes I)}{(I \otimes a)})=(\sapp{\sdag{a}}{I})\otimes (\sapp{I}{a})=\sapp{\sdag{a}}{a}$.
Note that we omit the type annotations $\quan{F}{\pmx}{t(2^{2})}$ in the term for simplicity.
By using the above transformation function, we can transform the system to a $t(2)$ typed one, named $\hat{H}_T$, as:

{\footnotesize
\begin{center}
$
\begin{array}{c@{\;\;}l}
&
\;
(\sdag{a}\textcolor{spec}{(0,0)}:\quan{F}{\pmx}{t(2)} \otimes {a}\textcolor{spec}{(0,1)}:\quan{F}{\pmx}{t(2)})\otimes (a\textcolor{spec}{(1,0)}:\quan{F}{\pmx}{t(2)} \otimes \sdag{a}\textcolor{spec}{(1,1)}:\quan{F}{\pmx}{t(2)})
\\[0.2em]
&
\quad
+(a\textcolor{spec}{(0,0)}:\quan{F}{\pmx}{t(2)} \otimes \sdag{a}\textcolor{spec}{(0,1)}:\quan{F}{\pmx}{t(2)})
\otimes
(\sdag{a}\textcolor{spec}{(1,0)}:\quan{F}{\pmx}{t(2)} \otimes {a}\textcolor{spec}{(1,1)}:\quan{F}{\pmx}{t(2)})
\\[0.3em]
=&
\frac{1}{2}(X + i Y) \otimes \frac{1}{2}(X - i Y) \otimes \frac{1}{2}(X - i Y) \otimes \frac{1}{2}(X + i Y)
+
\frac{1}{2}(X - i Y) \otimes \frac{1}{2}(X + i Y) \otimes \frac{1}{2}(X + i Y) \otimes \frac{1}{2}(X - i Y)
\\[0.3em]
=&
\frac{1}{8}(X\otimes X \otimes X \otimes X + X\otimes X \otimes Y \otimes Y+Y\otimes Y \otimes X \otimes X + Y\otimes Y \otimes Y \otimes Y
\\[0.1em]
&\quad\qquad+ X\otimes Y \otimes X \otimes Y + X\otimes Y \otimes Y \otimes X + Y\otimes X \otimes X \otimes Y + Y\otimes X \otimes Y \otimes X)
\end{array}
$
\end{center}
}

\begin{figure*}[t]
{\small
\begin{minipage}[b]{0.3\textwidth}
{\centering
{
  \Qcircuit @C=0.5em @R=0.5em {
    \lstick{} & \ctrl{1} & \qw          & \qw                            & \qw      & \ctrl{1} & \qw      \\
    \lstick{} & \targ    & \ctrl{1}     & \qw                            & \ctrl{1} & \targ & \qw \\
    \lstick{} & \qw      & \targ        & \multigate{1}{ZZ(2\theta)} & \targ    & \qw   & \qw \\
    \lstick{} & \qw      & \qw          & \ghost{{ZZ(2\theta)}}      & \qw      & \qw   & \qw
    }
}}
\subcaption{$\eexp{\sapp{ZZZZ}{\theta}}$ gate.}
\label{fig:zzzz-gate}  
\end{minipage}
\begin{minipage}[b]{0.33\textwidth}
{\centering
{
  \Qcircuit @C=0.5em @R=0.5em {
    \lstick{} & \gate{\cn{H}} & \multigate{3}{\eexp{\sapp{ZZZZ}{\theta}}} & \gate{\cn{H}} & \qw   \\
    \lstick{} &\gate{\cn{H}}  & \ghost{\eexp{\sapp{ZZZZ}{\theta}}}        & \gate{\cn{H}} & \qw  \\
    \lstick{} & \gate{\cn{H}} & \ghost{\eexp{\sapp{ZZZZ}{\theta}}}        & \gate{\cn{H}} & \qw  \\
    \lstick{} & \gate{\cn{H}} & \ghost{\eexp{\sapp{ZZZZ}{\theta}}}        & \gate{\cn{H}} & \qw
    }
}}
\subcaption{$\eexp{\sapp{XXXX}{\theta}}$ gate.}
\label{fig:xxxx-gate}  
\end{minipage}
\begin{minipage}[b]{0.33\textwidth}
{\centering
{
  \Qcircuit @C=0.5em @R=0.5em {
    \lstick{} & \qw & \multigate{3}{\eexp{\sapp{XXXX}{\theta}}} & \qw & \qw   \\
    \lstick{} &\qw  & \ghost{\eexp{\sapp{XXXX}{\theta}}}        & \qw & \qw  \\
    \lstick{} & \gate{\cn{S}} & \ghost{\eexp{\sapp{XXXX}{\theta}}}        & \gate{\sdag{\cn{S}}} & \qw  \\
    \lstick{} & \gate{\cn{S}} & \ghost{\eexp{\sapp{XXXX}{\theta}}}        & \gate{\sdag{\cn{S}}} & \qw
    }
}}
\subcaption{$\eexp{\sapp{XXYY}{\theta}}$ gate.}
\label{fig:xxyy-gate}  
\end{minipage}
}
\vspace*{-0.5em}
\caption{Circuits for quantum simulating the bosonic system.}
\label{fig:bosongates}
\vspace*{-1em}
\end{figure*}
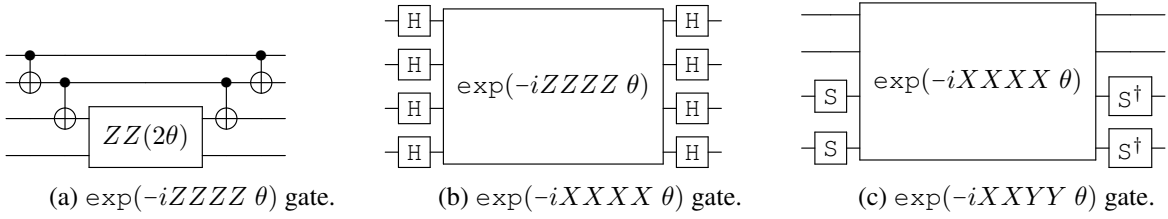

We apply the matrix exponent to the Hamiltonian to perform the simulation with respect to time $t$, as $\eexp{\sapp{\hat{H}_T}{t}}$.
By applying Trotterization, we can see that the simulation results in a sequence of unitary applications; each contains one of the tensor terms in the linear sum.
We use the two tensor terms $X\otimes X \otimes X \otimes X$ (abbreviated as $XXXX$) and $X\otimes X \otimes Y \otimes Y$ (abbreviated as $XXYY$) as an example.
Here, we want to generate the digital unitary circuits of $\eexp{\sapp{XXXX}{\theta}}$ and $\eexp{\sapp{XXYY}{\theta}}$ for some small period $\theta$, which results in the digital circuits in \Cref{fig:xxxx-gate,fig:xxyy-gate}, which are constructed based on the quadruple $Z$ interaction in \Cref{fig:zzzz-gate}.

As mentioned above, quantum gates are simulated based on target machine-level Hamiltonians, and simple quantum circuit generation might not be an effective program implementation.
There are several solutions. First, we can rewrite the quadruple $Z$ interaction for IBM machines to utilize more $ZZ$ or $XZ$ interactions built in the machines, such as in \Cref{fig:zzzz-gate}. 
Second, there are more effective bosonic system approximations \cite{peng2023quantum}, which can reduce the quadruple site interactions to two-site ones, such as $ZZ$ interactions, significantly optimizing the gate numbers in the simulation.
In fact, \cite{Cao_2015} approximated a higher-order site interaction with the linear sum of two or three site interactions.
In addition, other types of quantum machines support different forms of interactions, some of which allow more than two-site interactions \cite{Kyprianidis_2024}; such systems can better utilize quantum computers.

%% file: comparison.tex
\vspace{-0.5em}
\section{Related Work}
\label{sec:related}
\vspace{-0.5em}

We show some related works. Due to space limitations, we are unable to list many other fantastic works.

\noindent{\textbf{{Second Quantization Formalism.}}}
Second quantization, or occupation number representation, is a formalism used to describe and analyze quantum particle systems.
The key ideas of the methodology were introduced by Paul Dirac \cite{Dirac1988} and were later developed by Pascual Jordan \cite{Jordan1928} and Vladimir Fock \cite{Fock1932,Reed:1975uy}. 
In this approach, the quantum states are represented in the Fock state basis, which is constructed by filling up each single-particle state with a certain number of identical particles \cite{Becchi10}. Later, many works were developed for describing second quantization in analyzing different systems, such as fermions in lattice-based systems \cite{Levin:2003ws},
chemical and biological molecule systems \cite{Gori2023}, nuclear physical systems \cite{Johnson_2013,Christiansen2004}.
Recently, researchers developed algebraic formalisms based on second quantization and Lie algebra \cite{Batista_2004,SCHWARZ2021115601}
as a new extension of second quantization, in order to define the transformations from one particle system to another,
such as defining a generalization of Jordan-Wigner Transformation \cite{Batista_2001} to transform different particle systems to $t(2)$ typed particle system, a.k.a., quantum computer systems.
Ying \cite{ying2014quantum} developed a quantum version of recursion and program control flow based on creators and annihilators in second quantization.

\noindent{\textbf{{Quantum Digital-based Programming Languages.}}}
There are many digital quantum programming languages.
Q\# \cite{Svore_2018}, Quilc \cite{smith2020opensource}, ScaffCC \cite{JavadiAbhari_2015}, Project Q \cite{Steiger_2018}, Criq \cite{cirq_developers_2023_10247207}, Qiskit \cite{Qiskit2019} are industrial quantum digital circuit languages, without formal semantics, but based on a standardized denotational quantum digital circuit semantics.
There are formally verifying digital circuit programs, including  \qwire~\cite{RandThesis}, \sqir~\cite{PQPC}, and \qbricks~\cite{qbricks}, quantum Hoare logic and its subsequent works \cite{qhoare,qhoreusage,10.1145/3571222}, Qafny \cite{li2024qafny}. These tools have been used to verify a range of quantum algorithms, from Grover's search to quantum phase estimation.
There are works verifying digital circuit optimizations  (e.g., \voqc~\cite{VOQC}, CertiQ~\cite{Shi2019}), as well as verifying digital circuit compilation procedures,
including ReVerC~\cite{reverC} and ReQWIRE~\cite{Rand2018ReQWIRERA}.

There are functional language interpretations for quantum circuits \cite{10.1007/978-3-319-89366-2_19,DIAZCARO2019104012,Selinger2013,van_Tonder_2004,Altenkirch,10.1007/978-3-642-40922-6_5,Voichick_2023,Green2013,mingsheng-ying,qml-update,qml-thesis,10.1145/3571204}, with formal semantics.
These languages are designed to encode digital circuits as functional language constructs, such as quantum pattern matching.
Silq \cite{sliqlanguage} proposed a language to perform automatic uncomputation with denotational semantics for describing circuit behaviors.
ZX-calculus \cite{Coecke_2011,wang2023completeness} turns elements appearing quantum semantics in a diagram calculus for expressing unitary operations without measurement.

\noindent{\textbf{{Analog Quantum Computing Languages and Pulse Level Programming.}}}
There are a few Pulse-level and target machine Hamiltonian programming interfaces that view quantum computers as analog quantum simulators,
such as IBM Qiskit Pulse \cite{10.1145/3505636}, QuEra Bloqade \cite{bloqade2023quera}, and Pasqal
Pulser \cite{Silverio2022pulseropensource} developed by hardware service providers. 
There are computational quantum physics packages, e.g., QuTip \cite{JOHANSSON20121760}, supporting Hamiltonian simulation in a classical computer.
These works are inspired by many classical analog compilations \cite{10.1145/3373376.3378449,10.1145/2980983.2908116}
SimuQ \cite{10.1145/3632923} provides a formal language on analog quantum computing based on constructing Hamiltonian with Pauli groups.
Essentially, it models a $t(2)$ particle system based on a quantum computer setting.

\ignore{
\noindent{\textbf{\emph{Quantum and HPC computational Approaches for Analyzing Particle Systems.}}}
There are many software and algorithms proposed for analyzing particle systems.
Many works utilize HPC systems \cite{Zahariev2023,540141112,10.1063/5.0004997,6651029} to calculate molecule energy states based on some variational methods.
There are also software tools \cite{10.1145/3511715,10.1145/3503222.3507715,POWERS2021100696,PhysRevA.109.042418} for performing particle system Hamiltonian simulation, by viewing it as a quantum algorithm and compile it to quantum circuits.
Many other works \cite{Cervera_Lierta_2018,Yang_2020,YAMAGUCHI2002343,doi:10.1021/acs.jctc.2c00974} discuss simulating different particle systems in quantum computers.
Variational quantum eigensolver is one of the most well-known algorithms \cite{Tilly_2022,Cerezo2022} to analyze particle systems;
it is mainly used for performing energy state computation, such as estimating the ground energy state of water molecules \cite{Nam2020}.
}

%% file: conclusion.tex
\vspace{-1em}
\section{Conclusion}
\vspace{-0.5em}

We present \qsnd, aiming to define and analyze lattice-based particle systems.
We generalize quantum computer systems to be a specific quantum particle system, so we can view the compilation of quantum particle systems to quantum computers as mapping from one kind of particle system to another, definable in \qsnd.
Under the \qsnd generalization, we intend to connect the previously unlinked dots, fragmented in different layers, together, so one can utilize \qsnd to define the tasks in these dots, such as user applications, quantum particle system transformations, and target machine specifications.
%Moreover, we enable the definitions of many quantum computation models, such as circuit-based models, analog Hamiltonian simulations, and adiabatic evolution.
We provide typing information for different particle systems, which can be used to properly program a system and transform between different systems.
For future work, we intend to develop verification and testing toolchains based on \qsnd, as well as imposing different programming language mechanisms to help the analysis of \qsnd programs.
%For example, the exponential and logarithm operations can be analyzed by the least fixed point analysis tools, as well as different Hamiltonian analytical tools can be imposed to cut a large Hamiltonian into smaller ones that can be directly implemented in target machine Hamiltonians, such as the example in \Cref{sec:boson}.
%Finally, we plan to fully investigate the commutation and anti-commutation properties of bosons and fermions so that \qsnd can be used to analyze real-world particle systems.

%% file: background.tex
\section{Background}
\label{sec:background}

Here, we provide brief background information on quantum mechanics and quantum computing.
%A more comprehensive introduction is given in \Cref{appx:qmechan}.

\myparagraph{Second Quantization.}
Second quantization is a formalism to describe systems with varying numbers of particles. The creation (\(\sdag{a}\)) and annihilation (\(a\)) operators are used to add or remove particles from a quantum state, and more generally to express Hamiltonians.
Based on the formulations of the second quantization for lattice-based systems, the Hamiltonians portray the particles in a site as a vertex in a fixed lattice graph.
Many systems that are not inherently constructed as lattice-based systems can be analyzed based on a continuous version of the lattice-based one. 
\qsnd describes the second quantization for lattice-based systems.
 
\myparagraph{Quantum States.}\label{appx:state}
In the quantum computer setting, a quantum state $\varphi$ \footnote{In quantum mechanics, $\ket{\varphi}$ and $\bra{\varphi}$ are used to express a quantum state and a conjugate transpose of a state. We use $\varphi$ and $\sdag{\varphi}$ in this paper to avoid confusion with kets.} consists of one or more quantum bits (\emph{qubits}), which can be expressed as a two-dimensional Hilbert space vector $\begin{psmallmatrix} \alpha \\ \beta \end{psmallmatrix}$ where the \emph{amplitudes} $\alpha$ and $\beta$ are complex numbers; in a \emph{normalized qubit state}, $|\alpha|^2 + |\beta|^2 = 1$.
We frequently write the qubit vector as $\alpha\ket{0} + \beta\ket{1}$ (the Dirac notation \cite{Dirac1939ANN}), where $\ket{0} = \begin{psmallmatrix} 1 \\ 0 \end{psmallmatrix}$ and $\ket{1} = \begin{psmallmatrix} 0 \\ 1 \end{psmallmatrix}$ are \emph{computational kets}. When both $\alpha$ and $\beta$ are non-zero, we can think of the qubit being ``both 0 and 1 at once,'' a.k.a. in a \emph{superposition} \cite{mike-and-ike}, e.g., $\frac{1}{\sqrt{2}}(\ket{0} + \ket{1})$ represents an equal superposition of $\ket{0}$ and $\ket{1}$.
Larger quantum values can be formed by composing smaller ones with the \emph{tensor product} ($\otimes$) from linear algebra, e.g., the two-qubit value $\ket{0} \otimes \ket{1}$ (also written as $\ket{01}$) corresponds to vector $[~0~1~0~0~]^T$.
However, many multi-qubit values cannot be \emph{separated} and expressed as the tensor product of smaller values; such inseparable value states are called \emph{entangled}, e.g.\/ $\frac{1}{\sqrt{2}}(\ket{00} + \ket{11})$, known as a \emph{Bell pair}, which can be rewritten to $\Msum_{d=0}^{1}{\frac{1}{\sqrt{2}}}{\ket{dd}}$, where $dd$ is a bit string consisting of two bits, each of which must be the same value (i.e., $d=0$ or $d=1$). Each term $\frac{1}{\sqrt{2}}\ket{dd}$ is named a \emph{ket} \cite{mike-and-ike}, consisting an amplitude ($\frac{1}{\sqrt{2}}$) and a basis vector $\ket{dd}$.
In second quantization, a quantum state is typically expressed in a Fock space, a direct sum of different dimensional Hilbert spaces $\bigoplus_{j=0}^{\infty}\mathpzc{H}^{j}$,
for describing arbitrary particle movements with variable particle numbers;
to describe lattice-based particle systems, such formalism is usually expressed as tensors of Hilbert spaces (\Cref{sec:formal}).
%such formalism can be approximated by tensors of Hilbert spaces for lattice-based particle systems (\Cref{sec:formal}).
%such formalism should be understood as modeling possible particle movements (\Cref{sec:formal}). % can be described by arbitrary dimensional Hilbert spaces.
%People typically approximate the analysis of particle movements by tensors of Hilbert spaces; detailed descriptions are %in \Cref{sec:formal}. 

\myparagraph{Circuit-based Quantum Computation and Operations.}
Computation on a quantum value consists of a series of \emph{quantum operations}, each acting on a subset of particle sites in the quantum state. 
A quantum circuit model description expresses quantum computations as \emph{circuits}.
In these circuits, each horizontal wire represents a qubit, and boxes on these wires indicate quantum operations, or \emph{gates}. 
%For example, the circuit to create a Bell pair uses two qubits and applies two gates: the \emph{Hadamard} (\texttt{H}) gate and a \emph{controlled-not} (\texttt{CNOT}) gate. Applying a gate to a state evolves the state.
This evolution is mathematically expressed by multiplying the state vector by the gate's corresponding matrix representation; single-qubit gates are 2-by-2 matrices, and two-qubit gates are 4-by-4 matrices. The matrix representation of a gate must be \emph{unitary}, ensuring that it preserves the norm of the quantum state's amplitudes.
An entire quantum circuit can be described by composing its constituent gates into a single unitary matrix. This matrix representation encapsulates the overall transformation applied to the initial quantum state.

\begin{figure}[t]
\vspace*{-0.5em}
{\begin{center}
    \includegraphics[width=0.6\textwidth]{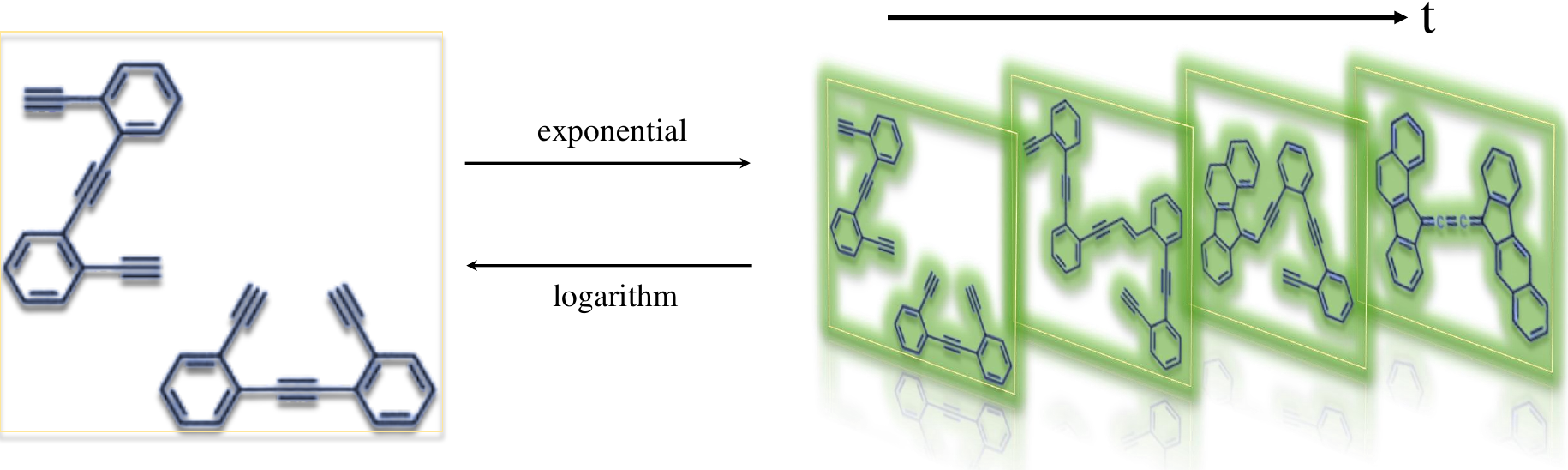}
 \end{center}
}
    \vspace*{-1em}
            \caption{Hamiltonian as a snapshot (left) and quantum simulation (unitary) as movie generation (right); $\cn{exp}$ turns a Hamiltonian to unitary and \cn{log} can perform a reverse.}\label{fig:ham-sim}   
\vspace*{-1.5em}
\end{figure}

\myparagraph{Two Modes and Two Applications in Second Quantization.}\label{appx:twoviews}
In second quantization, operations can be generalized to Hamiltonian $\hat{H}$, a Hermitian matrix ($\hat{H} = \sdag{\hat{H}}$), with two major applications, energy expectation value calculation, and quantum Hamiltonian simulation.
Second quantization for particle systems is how the scientific computing users in physics and chemistry think of programs; typically, they view a program having two modes, shown in \Cref{fig:ham-sim}, and the two modes corresponding to mainly two different applications below.

{\small
\begin{center}
$
E=\frac{
\sapp{\sdag{\varphi}}{(\sapp{\hat{H}}{\varphi})}
}
{\dabs{\varphi}^2}
\qquad
\qquad
\varphi(t) = \sapp{\eexp{\sapp{\hat{H}}{t}}}{\varphi(0)}
$
\end{center}
}
%\vspace*{-0.6em}

One can define a Hamiltonian matrix $\hat{H}$ to describe static particle interaction information as the snapshot mode, capturing a moment indicating the potential particle movements, analogized on the left of \Cref{fig:ham-sim}.
Defining the snapshot can help users compute the energy states of holding the particle structures in $\hat{H}$, the expectation value calculation (the left one above),
where ground energy state computation refers to finding the minimal expectation value among different choices of $\varphi$ (\Cref{sec:groundenergy}).
Additionally, one can perform Hamiltonian simulation, the simulation mode, by applying a matrix exponential to $\hat{H}$, as $\eexp{\sapp{\hat{H}}{t}}$ (can be understood as $\eexp{\sapp{t}{\hat{H}}}$), producing a unitary (executable in quantum computers). It is analogous to automatically generating a movie (right in \Cref{fig:ham-sim}), showing the trace particle movements propagating $\hat{H}$ in time $t$. Any circuit-based quantum computation can be interpreted as a form of Hamiltonian simulation.
One could also apply a logarithm to the unitary to get the intermediate snapshot for performing other applications.

Many applications can be formulated based on the two modes, e.g., adiabatic evolution \cite{farhi2000quantum} is a kind of Hamiltonian simulation, but its main purpose is to wait for the simulation to converge to a ground energy gate for eigenvalue computation.
In addition, many physical system users also intend to transform one particle system into another, analogous to compiling a program to another system for finding solutions.
One example is in \Cref{sec:qcompile}.

%where they define physical systems in terms of Hamiltonians with applications, such as performing ground energy state computation and Hamiltonian simulation. 
%Lattice-based Hamiltonian matrices (\Cref{sec:background}) describe particle interactions by locating particles in a fixed-site structure formed as a lattice, and Hamiltonian simulation simulates the Hamiltonian's transition behavior for a quantum state over a period of time.
%In this program view, the quantum computer behavior is a procedure of a lattice-based particle system Hamiltonian simulation.
%A trivial generalization is to generalize the Hamiltonian simulator view of quantum computers as a particle system, describable by second quantization, a standard physical formalism for describing quantum particle movements. 

\myparagraph{Quantum Measurement and Expectation Value Computation.}
In a quantum computer, a \emph{measurement} operation typically refers to computational basis measurement to extract classical information from a quantum state by collapsing the state to a basis state with a probability related to the value's amplitudes (\emph{measurement probability}), e.g., measuring $\frac{1}{\sqrt{2}}(\ket{0} + \ket{1})$ collapses the value to $\ket{0}$ with probability $\frac{1}{2}$,  and likewise for $\ket{1}$, returning classical value $0$ or $1$, respectively.
A more general Von Neumann measurement \cite{mike-and-ike} is to define a set of orthonormal basis $\{\varphi_i\}$, and define the measurement operation based on two normalized state $\varphi$ and $\varphi'$ ($\dabs{\varphi}=\dabs{\varphi'}=1$), as $\slen{\sapp{\sdag{\varphi}}{\varphi'}}^2$, where $\varphi$ is a linear sum of the above orthonormal bases, $\sdag{\varphi}$ is a row vector, and $\sapp{\sdag{\varphi}}{\varphi'}$ computes an inner product.
This means the following: given the state $\varphi'$, we measure the probability of projecting it to the state $\varphi$.
For example, we can define Hadamard bases $\ket{\pm}=\frac{1}{\sqrt{2}}(\ket{0}\pm\ket{1})$; the probabilities of projecting $\ket{+}$ on to $\ket{0}$ and the vice versa case, $\slen{\sapp{{\bra{0}}}{\ket{+}}}^2$ and $\slen{\sapp{{\bra{+}}}{\ket{0}}}^2$ ($\sdag{\ket{d}} =\bra{d}$), are both half.
In quantum computing, $\varphi$ can be achieved by applying a unitary $U$ to initial states $\ket{0}^{\otimes n}$.
if $\varphi=U\ket{0}^{\otimes n}$, we can think of the energy computation as  $\slen{\sapp{\sdag{\varphi}}{\varphi'}}^2=\slen{\sapp{\bra{0}^{\otimes n}}{(\sapp{\sdag{U}}{\varphi'})}}^2$,
i.e., we apply the inversed unitary of $U$ to $\varphi'$ and see the probability of measuring (computational basis measurement) out of zero.
The above expectation value calculation can be thought of as computing the inner product $\sapp{\sdag{\varphi}}{\varphi'}$, where $\varphi$ is normalized and $\varphi'$ is the result of $\sapp{\hat{H}}{\varphi}$. In the extended version of \qsnd, we do not directly include a measurement operation but include inner products and normalization operations to permit the definition of expectation value calculation, whereas measurement output operations, i.e., the operations that produce the probability of projecting a quantum state $\varphi$ onto $\varphi'$,  can be defined on top of these operations.

\ignore{
In summary, these foundational concepts of quantum mechanics—quantum states, operators, the Schrödinger equation, and Hilbert spaces—are crucial for understanding and developing quantum computing.

In quantum mechanics, when describing a quantum state, its basis element is typically not a qubit, but a particle site.
A quantum state is represented by a vector in a Hilbert space, which is a complete vector space with an inner product. For a single qubit, the state is described as:

\[
|\varphi\rangle = \alpha|0\rangle + \beta|1\rangle
\]

where \(|0\rangle\) and \(|1\rangle\) are basis states, and \(\alpha\) and \(\beta\) are complex coefficients such that \(|\alpha|^2 + |\beta|^2 = 1\). This representation is known as Dirac notation, or bra-ket notation \cite{dirac1939}. In this context, we will consider only normalized states, which satisfy \(\langle \varphi | \varphi \rangle = 1\).

%\liyi{below is subject to change. No reviewer in POPL will understand any of this. The problem is that they should not try hard to understand this. This is background, and the purpose is to make them comfortably accept these. }
\section*{Hilbert Spaces and Tensor Products}

The Hilbert space of a single qubit is a two-dimensional complex vector space. For multiple qubits, the total Hilbert space is the tensor product of the individual qubit spaces. For two qubits:

\[
\mathcal{H} = \mathcal{H}_1 \otimes \mathcal{H}_2
\]

where \(\mathcal{H}_1\) and \(\mathcal{H}_2\) are the Hilbert spaces of the individual qubits, each of dimension 2. The combined space has dimension \(2^2 = 4\). For \(n\) qubits, the total Hilbert space has dimension \(2^n\).

\section*{Operators}

Operators act on quantum states to produce new states or to extract information about the system. Key types of operators in quantum mechanics include Hermitian and unitary operators.

\subsection*{Hermitian Operators}

An operator \(\hat{A}\) is Hermitian if it equals its own conjugate transpose:

\[
\hat{A} = \hat{A}^\dagger
\]

Hermitian operators correspond to observables, quantities that can be measured. Their eigenvalues are real numbers, and their eigenvectors form an orthonormal basis for the Hilbert space. Examples of Hermitian operators include:

\liyi{bad example. no CS people will understand.}
\begin{itemize}
    \item \textbf{Spin operator (\(\hat{S}_z\))}: Measures the spin of a particle along the z-axis.
    \item \textbf{Hamiltonian (\(\hat{H}\))}: Represents the total energy of the system.
\end{itemize}

The expectation value of a Hermitian operator \(\hat{A}\) in a state \(|\varphi\rangle\) is given by:

\[
\langle \hat{A} \rangle = \langle \varphi | \hat{A} | \varphi \rangle
\]

This represents the average measured result of the physical variable corresponding to \(\hat{A}\) when the system is in state \(|\varphi\rangle\).

\subsection*{Unitary Operators}

An operator \(\hat{U}\) is unitary if it satisfies:

\[
\hat{U}^\dagger \hat{U} = \hat{U} \hat{U}^\dagger = \hat{I}
\]

Unitary operators preserve the inner product and hence the norm of quantum states. They are used to describe the evolution of isolated quantum systems and perform quantum gates in quantum computing. Notably, the exponential of a Hermitian operator \(\hat{A}\) is unitary:

\[
\hat{U} = e^{i\hat{A}}
\]

Examples of unitary operators include:

\begin{itemize}
    \item \textbf{Rotation operator (\(\hat{R}(\theta)\))}: Rotates the state vector in a plane by an angle \(\theta\).
    \item \textbf{Pauli-X gate}: Acts like a classical NOT gate, flipping the state \(|0\rangle\) to \(|1\rangle\) and vice versa.
    \item \textbf{Hadamard gate}: Creates superposition by transforming \(|0\rangle\) into \(\frac{1}{\sqrt{2}}(|0\rangle + |1\rangle)\) and \(|1\rangle\) into \(\frac{1}{\sqrt{2}}(|0\rangle - |1\rangle)\).
\end{itemize}

\section*{The Schrödinger Equation}

The dynamics of a quantum system are governed by the Schrödinger equation.

\subsection*{Time-Independent Schrödinger Equation}

\[
\hat{H}|\varphi\rangle = E|\varphi\rangle
\]

where \(\hat{H}\) is the Hamiltonian, \(|\varphi\rangle\) is an eigenstate, and \(E\) is the energy eigenvalue. The ground state corresponds to the lowest energy eigenvalue, while excited states have higher energies.

\subsection*{Variational Principle}

To find the exact ground state energy \(E_0\), one can use the variational principle. For a normalized state \(|\phi\rangle\), the expectation value of the Hamiltonian over all possible states \(|\varphi\rangle\) is minimized:

\[
E_0 = \min_{|\varphi\rangle} \langle \varphi|\hat{H}|\varphi\rangle
\]

By minimizing this expectation value over all possible normalized states, the exact ground state can be obtained.

\subsection*{Time-Dependent Schrödinger Equation}

\[
i \frac{\partial}{\partial t} |\varphi(t)\rangle = \hat{H}|\varphi(t)\rangle
\]

where \(t\) is the time variable. Its solution involves the propagator \(e^{-i\hat{H}t}\), which evolves the state from an initial state \(|\varphi(0)\rangle\):

\[
|\varphi(t)\rangle = e^{-i\hat{H}t}|\varphi(0)\rangle
\]

\noindent\textbf{\textit{Quantum Computations.}}

From the perspective of quantum mechanics, quantum computation can be viewed as a quantum simulation procedure based on the Schrödinger equation, involving unitary operations and quantum states. In this context, a quantum unitary operation may be defined as an exponential of a Hamiltonian.

Typical quantum computing theory often omits the explicit Hamiltonian and focuses on the computational aspects, treating the evolution of a quantum state as a sequence of \emph{quantum operations}, each acting on a subset of qubits in the quantum state. In standard presentations, quantum computations are expressed as \emph{circuits}. In these circuits, each horizontal wire represents a qubit, and boxes on these wires indicate quantum operations, or \emph{gates}. 

For example, the circuit to create a Bell pair uses two qubits and applies two gates: the \emph{Hadamard} (\texttt{H}) gate and a \emph{controlled-not} (\texttt{CNOT}) gate. Applying a gate to a state evolves the state. This evolution is mathematically expressed by multiplying the state vector by the gate's corresponding matrix representation; single-qubit gates are 2-by-2 matrices, and two-qubit gates are 4-by-4 matrices. The matrix representation of a gate must be \emph{unitary}, ensuring that it preserves the norm of the quantum state's amplitudes.

An entire quantum circuit can be described by composing its constituent gates into a single unitary matrix. This matrix representation encapsulates the overall transformation applied to the initial quantum state.

In our approach, we provide a time-dependent quantum mechanics perspective on quantum computation by incorporating both the Hamiltonian and time-propagation based on the time-dependent Schrodinger equation.

\noindent\textbf{\textit{No Cloning and Observer Effect.}} 
The \emph{no-cloning theorem} suggests no general way of copying a
quantum value. In quantum circuits, this is related to ensuring the reversible
property of unitary gate applications.
For example, the Boolean guard and body of a quantum conditional cannot refer to
the same qubits, e.g., $\qif{x[1]}{\qass{x[1]}{x[1]\, \splus\, 1}}$ violates the property as
$x[1]$ is mentioned in the guard and body.
The quantum \emph{observer effect} refers to leaking information from a quantum value state. If a quantum conditional body contains a measurement or classical variable updates, it causes the quantum system to break down due to the observer effect. 
\qafny{} enforces no cloning and no observer breakdown through the syntax and flow-sensitive type system.
}

%% file: appendix.tex
%\appendix

\section{Energy Expectation Value Calculation.}\label{sec:groundenergy}

Other than Hamiltonian simulation, another key application is the energy state computation based on expectation value calculation, described above.
One of the central pieces in energy state computation is to find the exact ground state energy expectation value \(E_0\) and the state $\varphi$ holding the energy value.
Based on the variational principle, the expectation value of the Hamiltonian over all possible states $\varphi$ is minimized:

%\vspace{-0.3em}
{
\begin{center}
$
{\displaystyle E_0 = \min_{\varphi} \sapp{\sdag{\varphi}}{(\sapp{\hat{H}}{\varphi}})}
$
\end{center}
}
%\vspace{-0.3em}

The exact ground state can be obtained by minimizing this expectation value over all possible normalized states.
In \qsnd, we compile the computation to HPC software through classical optimization techniques, such as gradient descent methods, to ``guess'' the minimal expectation value and its associated state of a given Hamiltonian.
Many HPC particle computation platforms use other sophisticated techniques based on the guessing-out strategy. For example, Gamess \cite{540141112,Zahariev2023} and NWChem \cite{10.1063/5.0004997} are HPC software to compute energy states of chemistry molecule structures, and users can define the structures in second quantization in the software.
The quantum supremacy of the energy state calculation problems is murky, as some researchers \cite{Lee2023} do not believe in the existence of such supremacy.
Nevertheless, the typical way to calculate in quantum computers is through the variational quantum eigensolver concept \cite{TILLY20221}, having similar procedures as the HPC counterpart.

\section{Linear Sums and Quantum Choices}\label{appx:linearsum}

A linear sum in a state connects kets in \Cref{fig:data} represents superposition state; the sum operation \(e_1 + e_2\) is reinterpreted a \emph{quantum choice} operation, in the sense that one can perform either \(e_1\) or \(e_2\) to the state \(\varphi\) at the same time,
e.g., it might produce a superposition if \(\varphi\) is a state just initialized as a ket.

When second quantization users use quantum choices to model a lattice-based molecule system, they usually have two intentions.
The \emph{first intention} is 
%similar to the choice selection in the equal-sum example above, and we intend 
to "create" a state going in two different directions simultaneously in a possible superposition.
For example, applying a quantum choice operation $a + \sdag{a}$ on a $t(4)$ particle state $\ket{1}$, with state \emph{normalization}, results in the following:

{\small
\begin{center}
$\frac{1}{\sqrt{2}}\ket{0}+\frac{1}{\sqrt{2}}\ket{2}$
\end{center}
}

In the Hubbard system $\hat{H}$, for each $j$-th site, the sum of the two terms, $\sdag{a}(j) a(j\splus 1)$ and $\sdag{a}(j\splus 1) a(j)$, model all the possibilities of electrons, being relocated to adjacent particle site. 
Normalization means that the norm of the amplitudes in different kets must be summed to $1$, referring to the total probabilities of different parts.
State normalization permits Hamiltonian simulation and quantum computing, while the non-normalized state plays a role in computing expectation values.
In the extended version of \qsnd, we permit a \cn{nor} operation to normalize a state, and quantum simulation is essentially defined as $\sapp{\eexp{\sapp{\hat{H}}{t}}}{\cn{nor}(\varphi(0))}$ for arbitrary state $\varphi(0)$, applying $\eexp{\sapp{\hat{H}}{t}}$ to a normalized state $\cn{nor}(\varphi(0))$.
We show a theorem in \Cref{sec:theorem} to link simulations on normalized and non-normalized states.

The \emph{second intention} is to add repulsion, i.e., we view a quantum choice term as some extra force applied to the system to increase or decrease the probability of observing a basis vector state.
Recall that the normalization procedure ensures that a system's total probability is $1$. Conceptually, adding a quantum choice term for certain basis vectors with a positive or negative amplitude increases or reduces the choice of observing the basis vectors, respectively.
For example, the term $z_u \sum_{j} \sdag{a}(j) a(j) \sdag{a}(j\splus 1) a(j\splus 1)$, in the above Hubbard system, usually refers to the repulsion force, reducing the likelihood of the happening of some basis vectors. Let's name the above term as $\hat{H}_u$ and the term $\sminus z_t \sum_{j} (\sapp{\sdag{a}(j)}{ a(j\splus 2)} + \sapp{\sdag{a}(j\splus 2)}{ a(j)})$ as $\hat{H}_t$.
Notice that the complex amplitudes $\sminus z_t$ and $z_u$ have opposite signs.
Applying the whole Hamiltonian to a quantum state $\varphi$ results in a quantum state of two pieces: $(\sapp{\hat{H}_t}{\varphi})+(\sapp{\hat{H}_u}{\varphi})$, connected by $+$. A non-zero ket in the state $\sapp{\hat{H}_u}{\varphi}$ has an amplitude with an opposite sign as the same basis-vector amplitude in $\sapp{\hat{H}_t}{\varphi}$, which means that the sum of their amplitudes is reduced; thus, the quantum choice term $\hat{H}_u$ acts as a penalty for a specific group of ket states.  
To see how this happens, realize that the Hubbard system describes $t(2)$ typed particles, so there are four cases of the basis vectors of two adjacent particles:

{\small
\begin{center}
$\dualb{}{0}{0} \qquad\qquad\dualb{}{0}{1} \qquad\qquad \dualb{}{1}{0} \qquad\qquad \dualb{}{1}{1}$
\end{center}
}

Assuming that we are dealing with a single particle site, the $\hat{H}_u$ term turns all basis vectors, except $\dualb{}{1}{1}$, to $\zero$, where $\dualb{}{1}{1}$ actually represents two electrons in opposite spins.
Assume that a single $t(2,2)$ typed particle state $\varphi$ contains the ket $\dualb{z}{1}{1}$, applying $\hat{H}_u$ to a state $\varphi$ results in a ket $\dualb{z_u * z}{1}{1}$.
When applying $\hat{H}_t$ to $\varphi$, we expect to have another ket $\dualb{\sminus z_t * z}{1}{1}$, with the same basis vector as above.
The sum of the two sub-parts turns the amplitude of the basis vector $\dualb{}{1}{1}$ to become $z*(z_u-z_t)$, with $\slen{z*(z_u-z_t)}\le\slen{z}$.
This is why the additional repulsion term acts as a penalty on the ket $\dualb{}{1}{1}$ to reduce its probability of happening.

\section{Gate Syntheses}\label{appx:gatesyn}

At the quantum computer machine level, every gate is implemented as the simulation of a gate Hamiltonian over a certain time period. To define a unitary gate \(U\), theoretically, we need to find its snapshot Hamiltonian by taking the logarithm $\elog{U}$.
Since $U$ is usually small, we usually perform reverse engineering by computing the eigendecomposition of \(U\) and forming a $t(2)$ typed Hamiltonian \(\hat{H}_U\) based on the eigenstates found through eigendecomposition, and simulates the gate based on \(\eexp{\hat{H}_U t}\) with respect to time $t$. We show below the simulation of a Hadamard (\cn{H}) and controlled-not (\cn{CX}) gate, with respect to time period \(\pi\).

{\small
\[
\hat{H}_{\cn{H}} = (\frac{1}{2} \sminus \frac{\sqrt{2}}{4}) \sapp{a}{\sdag{a}} + (\frac{1}{2} \sminus \frac{\sqrt{2}}{4} ) \sapp{\sdag{a}}{a} + (\sminus \frac{\sqrt{2}}{4})X
\qquad
\cn{H} = \eexp{\hat{H}_{\cn{H}} \pi}
\]
}

The above formula is the one particle Hamiltonian for simulating a Hadamard gate (\cn{H}). To reverse engineer the Hamiltonian, we perform an eigendecomposition on Hadamard gates (\cn{H}), find an eigenstate \(\varphi = -\sin(\frac{\pi}{8})\ket{0} + \cos(\frac{\pi}{8})\ket{1}\). We can then construct Hadamard Hamiltonians by the matrix multiplication $\sapp{\varphi}{\sdag{\varphi}}$, which is the \qsnd Hamiltonian $\hat{H}_{\cn{H}}$ above. The Hadamard gate (\cn{H}) can then be defined as a simulation of the Hamiltonian, as \(\cn{H} = \eexp{\hat{H}_{\cn{H}} \pi}\), with the time period \(\pi\).
Similarly, we can find the two-particle Hamiltonian for a controlled-not gate (\cn{CX}) through eigendecomposition and implement it in \qsnd, as shown below. The gate can then be defined as a simulation of the Hamiltonian as \(\cn{CX} = \eexp{\hat{H}_{\cn{CX}} \pi}\), with the time period \(\pi\).

{\small
\[
\hat{H}_{\cn{CX}} = \sapp{a}{\sdag{a}} \otimes (\sminus \frac{1}{2} X + \frac{1}{2} I)
\qquad
\cn{CX} = \eexp{\hat{H}_{\cn{CX}} \pi}
\]
}

In short, matrix exponential operations turn a Hamiltonian snapshot into a quantum system simulation, while a matrix logarithm operation, modeled in \qsnd, reverses a simulation to a Hamiltonian snapshot. For example, by applying a matrix logarithm on the \cn{H} and \cn{CX} gates, we might find a set of results, and \(\hat{H}_{\cn{H}}\) and \(\hat{H}_{\cn{CX}}\) are in the set \footnote{Matrix logarithms are not unique. \qsnd fixes the uniqueness by choosing one implementation. However, the \qsnd type system and equivalence relations are general enough without relying on the specific implementation.}.

\section{Using \qsnd to Tackle Fermion Systems}\label{sec:fermions}

Here, we show how to extend \qsnd to support the analysis of the fermion system, starting with the extension of \qsnd's types and semantics.

\subsection{Modeling Fermion Particles}

\begin{figure*}[h]
{\small

  \[
        \text{Fermion Flag}\qquad \aleph\qquad
        \begin{array}{l@{\quad}l@{\;\;}c@{\;\;}l@{\;\;}} 
       \text{Expression} & e & ::= &v:t^{[\aleph]}(m)\mid\alpha:\quan{F}{\zeta}{t^{[\aleph]}(m)}\mid ...
    \end{array}
  \]
}

{\small
  \begin{mathpar}
    \inferrule[]{}
  {t^{\aleph}(2) \simeq \aleph}
               
    \inferrule[S-TenF]{v' = \sapp{\alpha^{[\dag]}:\quan{F}{\zeta}{\iota}}{v}}
                { \teq{\sapp{(\alpha^{[\dag]}:\quan{F}{\zeta}{\iota} \otimes e)}{(v \otimes e')}}{v' \otimes (\sapp{\textcolor{spec}{\funsa{S}{\iota}{v'}}}{\sapp{e}{e'}})}}
\end{mathpar}
}
{\small
\begin{center}
$
\textcolor{spec}
{
\funsa{S}{\aleph(n)}{\Motimes_{k}\ket{j_k}} = (\sminus 1)^{\sum_{k} j_k}
\qquad
\funsa{S}{t(n,m)}{\Motimes_{k}\ket{j_k}} = 1
}
$
\end{center}
}
\caption{Extending \qsnd Formalism to Fermion Systems. $[\aleph]$ means either having or not having $\aleph$.}
\label{fig:fermion}
\end{figure*}

To extend \qsnd for describing fermion particle behaviors, we extend the quantum state type to include a flag $\aleph$, where $t^{\aleph}(m)$ type refers to a fermion particle state, and $m=2$ for any fermion, meaning that every basis-vector only has two states. The anti-commutation property guarantees that every fermion is a two-state system.
Therefore, we can abbreviate a fermion type $t^{\aleph}(2)$ as $\aleph$.
The biggest difference between fermions and boson-like particles is the anti-commutation property of fermion quantum states.
The concept is that any quantum fermion state $\varphi$ can be rewritten to the application of a second quantization formula, containing creators and annihilators, to a $\ket{0}^{\otimes n}$ state.
Then, in the second quantization formula, if there are two $\alpha_1$ and $\alpha_2$ where $\alpha_1\neq\alpha_2$, then $\sapp{\sapp{\alpha_1}{\alpha_2}}{(v:\aleph(n))}=(\sminus 1)\sapp{\sapp{\alpha_2}{\alpha_1}}{(v:\aleph(n))}$; here $\alpha$ refers to $\sdag{a}$ or $a$.

To enforce fermion anti-commutation, we maintain a quantum state in a certain structure by multiplying the correct $\sminus 1$ by a ket.
For a quantum state $\varphi$, with many particle sites, each of which contains many spins, we can impose an order to every site in the while state,
e.g., a state is arranged as $w_0 \otimes ... \otimes w_n$, where each $w_k$ is a ket, and a superposition quantum state can be expanded through the $+$ operation.
To apply a matrix operation to $v_j$, we count the number ($m$) of $\ket{1}$ vectors for all $v_k$, such that $k < j\sminus 1$, and apply $(\sminus 1)^{m}$ to the state.

To enforce the above counting mechanism, we rewrite the semantic rule \rulelab{S-TenF} to include a function $\mathpzc{S}$.
The functional implementation of $\mathpzc{S}$ shows the polymorphism of the \qsnd system and maintains special commutation properties for different kinds of particles, such as boson-like particles and fermions, depending on the types of the particles.
For boson-like particles (type $t(m)$), the function produces an identity $1$, while for fermions (type $\aleph$), the $\mathpzc{S}$ function in \Cref{fig:fermion} performs the above counting mechanism.
Here, we also need to track the right position of placing $j$, i.e., imposing the order we manage on the quantum state, to ensure the right counts can be achieved.
To do so, we use rule \rulelab{S-TenF} to track the right positions for placing a $\mathpzc{S}$ function with the right counts.

{\small
\begin{center}
$
\sapp{a(j)}{z_0\ket{m_0} \otimes ... \otimes z_j\ket{m_j} \otimes ...\otimes z_n\ket{m_n}} 
\equiv 
\textcolor{spec}{(-1)^{\sum_{k=0}^{j\sminus 1}m_k}}{z_0\ket{m_0} \otimes ... \otimes z_j (\sapp{a}{\ket{m_j}}) \otimes ...\otimes z_n\ket{m_n}} 
$
\end{center}
}

In the above example, with all sites being $\aleph$ type, the application $a(j)$ results in the blue part amplitude sign above, where we compute $\sum_{k=0}^{j\sminus 1}m_k$.
The functional implementation of $\mathpzc{S}$ permits the polymorphism of the \qsnd system, depending on the types of quantum states, i.e., when the type is $t(m)$, $\mathpzc{S}$ outputs an identity $1$.
%To properly type fermion particles, we need to include the type rule \rulelab{T-Fem} to ensure that every spin of a $\aleph(n)$ typed ket is within the range $[0,1]$. 

\subsection{Jordan-Wigner Transformation for Fermions}\label{sec:jw-trans}

The above section discusses the language features of having fermions.
To analyze fermions in a quantum computer, we need to transform the $\aleph$ typed particle system to a $t(2)$ typed system; such compilation is done by Jordan-Wigner transformation.
The Jordan-Wigner transformation is a critical tool in quantum mechanics and condensed matter physics. It allows for the mapping between Pauli systems ($X$, $Y$, $Z$) and fermionic/bosonic systems. This transformation is particularly useful in the study of lattice models, as it enables the analysis of spin chains in the language of fermions and facilitates the application of methods and insights from fermionic systems to solve spin-based problems.
The Jordan-Wigner transformation is a mapping that allows the transformation of spin operators into fermionic creation and annihilation operators, while the inverse equations are useful to transform creators and annihilators into quantum computers based on Pauli systems. 

%The Jordan-Wigner transformation provides a way to convert between spin operators and fermionic creation and annihilation operators. This is crucial in one-dimensional lattice models, enabling the transformation of spin models, like the XY model or the transverse field Ising model, into models of spinless fermions.
%The Jordan-Wigner transformation is a mapping that allows the transformation of spin operators into fermionic creation and annihilation operators. This transformation is particularly useful in the study of one-dimensional spin systems and quantum chains.

\subsubsection*{Jordan-Wigner Transformation for a single lattice site}

The Pauli matrices \(X\), \(Y\), and \(Z\) form the basis for the spin-$\frac{1}{2}$ operators. These matrices satisfy the commutation and anti-commutation relations.

The identity matrix \(\mathbb{I}\) and the zero matrix ($\zero$, matrix format: $\begin{psmallmatrix} 0 & 0 \\ 0 & 0 \end{psmallmatrix}$) are used to represent the unit operator and the null operator, respectively. We now define the anti-commutation relation of the Pauli matrices:

\[
\{M, N\} = 2 \sapp{\delta^{M N}}{I}, \quad M, N \in \{X, Y, Z\} 
\] 

Here, \(\delta^{M N}\) is the Kronecker delta, which is $1$ when $M$ and $N$ coincide and $0$ otherwise, indicates that the square of each Pauli matrix is the identity matrix
$M^2 = I$. In describing the Jordan-Wigner Transformation, much literature describes the raising and lowering operators, which are essentially the $t(2)$ typed creators and annihilators, and their relations with respect to the Pauli operations are given in \Cref{sec:boson}, as we restate below.

{
\[
\sdag{a}:\quan{F}{\pmx}{t(2)}= \frac{1}{2}\left(X + i Y\right)
\qquad
a:\quan{F}{\pmx}{t(2)}= \frac{1}{2}\left(X - i Y \right)
\]
}

The inverse equations above can be used to define $t(2)$ typed Pauli bases $X$, $Y$, and $Z$.

{
\[
\begin{array}{l}
X = \sdag{a}:\quan{F}{\pmx}{t(2)}+ a:\quan{F}{\pmx}{t(2)}
\\ Y = i(a:\quan{F}{\pmx}{t(2)} - \sdag{a}:\quan{F}{\pmx}{t(2)})
\\ Z = 2\sapp{\sdag{a}:\quan{F}{\pmx}{t(2)}}{a:\quan{F}{\pmx}{t(2)}} - I
\end{array}
\]
}

\myparagraph{Jordan-Wigner Transformation for many Particle Sites}
As we mentioned in \Cref{fig:data}, fermion with many particles might have different compilation scheme, due to enforcing anti-commutation.
Compiling a two particle fermion is to compile a $\aleph \otimes \aleph$ typed system to a $t(2)$ typed Pauli system.
We show below the compilation of such a typed system. Here, we use $\alpha_0$ to refer to applying $\alpha$ to the first site and $\alpha_1$ to refer to applying $\alpha$ to the second site, with $\alpha \in \{a, \sdag{a}\}$.

\[
\sdag{a_0}:\quan{F}{\pmx}{\aleph \otimes \aleph} = \sdag{a}:\quan{F}{\pmx}{t(2)} \otimes I \qquad a_0:\quan{F}{\pmx}{\aleph \otimes \aleph} = a:\quan{F}{\pmx}{t(2)} \otimes I
\]

\[
\sdag{a_1}:\quan{F}{\pmx}{\aleph \otimes \aleph} = Z \otimes \sdag{a}:\quan{F}{\pmx}{t(2)} \qquad a_1:\quan{F}{\pmx}{\aleph \otimes \aleph} = Z \otimes a:\quan{F}{\pmx}{t(2)}
\]

The above compilation results satisfy the anti-commutation relations expected of fermionic operators.

\[
\{{a_0}:\quan{F}{\pmx}{\aleph \otimes \aleph}, \sdag{a_0}:\quan{F}{\pmx}{\aleph \otimes \aleph}\} = I \qquad \{{a_1}:\quan{F}{\pmx}{\aleph \otimes \aleph}, \sdag{a_1}:\quan{F}{\pmx}{\aleph \otimes \aleph}\} = I
\]

In addition, they commute with each other:

\[
\{\sdag{a_0}:\quan{F}{\pmx}{\aleph \otimes \aleph}, {a_1}:\quan{F}{\pmx}{\aleph \otimes \aleph}\} = \zero, \quad \{\sdag{a_1}:\quan{F}{\pmx}{\aleph \otimes \aleph}, {a_0}:\quan{F}{\pmx}{\aleph \otimes \aleph}\} = \zero
\]

The above compilation corresponds to the use of the $\mathpzc{S}$ function in our semantics to guarantee the anti-commutation of fermions above, where we first fix the order of particle sites in a one-dimensional ($1D$) plane, and then we multiply $-1$ to a ket state depending on the counts of the number of occupied spins.
In Pauli group representation, multiplication refers to adding $Z$ terms.
For systems with more than two spins, the transformation extends to include additional spins.

\[
\sdag{a_0}:\quan{F}{\pmx}{\otimes^3 \aleph} = \sdag{a}:\quan{F}{\pmx}{t(2)} \otimes I \otimes I \qquad a_0:\quan{F}{\pmx}{\otimes^3 \aleph} = a:\quan{F}{\pmx}{t(2)} \otimes I \otimes I
\]

\[
\sdag{a_1}:\quan{F}{\pmx}{\otimes^3 \aleph} = Z \otimes \sdag{a}:\quan{F}{\pmx}{t(2)} \otimes I \qquad a_1:\quan{F}{\pmx}{\otimes^3 \aleph} = Z \otimes a:\quan{F}{\pmx}{t(2)} \otimes I
\]

\[
\sdag{a_2}:\quan{F}{\pmx}{\otimes^3 \aleph} = Z \otimes Z \otimes \sdag{a}:\quan{F}{\pmx}{t(2)} \qquad a_2:\quan{F}{\pmx}{\otimes^3 \aleph} = Z \otimes Z \otimes a:\quan{F}{\pmx}{t(2)}
\]

This extension allows the representation of quantum particle interactions in longer quantum chains.
This compilation pattern is general to define the case for $n$ particle site, e.g.,
to compile a $\aleph$ system with $n$ particle sites, we perform the following transformation $\gg^n$:

{\small
\[
    a_j:\quan{F}{\pmx}{\otimes^3 \aleph} \gg^n  \Motimes_{k=0}^{j\sminus 1} Z  \otimes a:\quan{F}{\pmx}{t(2)} \otimes \Motimes_{k=j\splus 1}^{n\sminus 1} \qquad
    \sdag{a_j}:\quan{F}{\pmx}{\otimes^3 \aleph} \gg^n \Motimes_{k=0}^{j\sminus 1} Z  \otimes \sdag{a}:\quan{F}{\pmx}{t(2)} \otimes \Motimes_{k=j\splus 1}^{n\sminus 1} 
\]
}

Based on the above analysis, to define the Jordan-Wigner Transformation for an $m$ fermion system as a function $\gg$ in \qsnd, as $\gg:\quan{F}{\zeta}{\Motimes^m \aleph} \to \quan{F}{\zeta}{\Motimes^m t(2)}$, the function is the function $\gg^n$.

\section{A Hubbard Model for Hydrogen Chains}\label{sec:hubbard} % Added a label. Feel free to change/remove

The Hubbard model (system) describes the interactions between elementary particles, specifically focusing on the electrons having fermion behaviors. 
Here, we focus on the one-dimensional Hydrogen chain, one of the most quintessential systems described using the Hubbard system.
In this system, we can view a site as a $\Motimes^2 \aleph$ typed Hydrogen atom (fermion). \Cref{eq:hubbard} describes the Hamiltonian for this system.

There are two terms in the Hamiltonian. The first accounts for the energy due to the movement (hopping) of electrons, whereas the second term accumulates the energy due to electron repulsion. As the atoms are arranged in a one-dimensional array, the electrons can only move (hop) to the neighboring adjacent atoms, such as moving from $j$-th to $j\splus 1$-th site, in this system.

In simulating the Hubbard system, users typically try to manipulate different $z_t$ and $z_u$ values for using the Hubbard system to analyze different particle behaviors.
In simulating the Hydrogen chain  \cite{melo_2021}, we assign a constant to $z_t$ and make $z_u$ depending on the time periods.
For a period of $T$. $z_u$ will transit from the initial value $z_{u0}$ at time $t = 0$ to the final value $Z_{uf}$ at time $t = T$,
and the equation looks like $ z_u(t) = (1 - \frac{t}{T} )z_{u0} + \frac{t}{T}z_{uf}$.
The naive compilation of the Hubbard system is similar to the one described in \Cref{sec:boson}, except that we also need to include $Z$ terms by enforcing anti-commutation.
To better compile the system, previous researchers \cite{melo_2021} tried to use unconventional quantum state mapping from a $\Motimes^2 \aleph$ typed state to a $\Motimes^2 t(2)$ typed qubit state. For example, for a two-particle system, they utilize an optimized state mapping as follows, where we mark $\ket{k}_0$ and $\ket{k}_1$ as the first and second quantum particle, respectively.

{\small
\begin{center}
$
\begin{array}{l}
\ket{1}_0\ket{1}_1\otimes \ket{0}_0\ket{0}_1\to \ket{0}\ket{0}
\qquad
\ket{1}_0\ket{0}_1\otimes \ket{0}_0\ket{1}_1\to \ket{0}\ket{1}
\\[0.1em]
\ket{0}_0\ket{1}_1\otimes \ket{1}_0\ket{0}_1\to \ket{1}\ket{0}
\qquad
\ket{0}_0\ket{0}_1\otimes \ket{1}_0\ket{1}_1\to \ket{1}\ket{1}
\end{array}
$
\end{center}
}

They assume that the other ket states do not exist.\
With the unconventional mapping, they can rewrite the Hubbard system to an Ising system as:

{\small
\begin{center}
$
\hat{H}_S=\sminus z_t(X \otimes I + I \otimes X) + z_u Z \otimes Z
$
\end{center}
}

The compilation of this new optimized system is similar to the one in \Cref{sec:qcompile}, i.e., through Trotterization, the simulation of the system generates a series of $X$-axis rotation gate $\cn{Rx}$ as well as $\cn{ZZ}$ interaction gates.

\myparagraph{Determining the Parameters \(z_t\) and \(z_u\) in Hubbard System.}

The parameters \(z_t\) and \(z_u\) in the Hubbard system are crucial for accurately describing the physical system. For a $1D$ chain of hydrogen atoms, these parameters can be determined as follows:

1. \textbf{Hopping Integral \(z_t\)}: The hopping integral \(z_t\) represents the kinetic energy associated with an electron hopping from one site to another. It can be estimated using the overlap integral of the atomic orbitals on neighboring sites. For hydrogen atoms, the 1s orbitals are used. The hopping integral can be calculated as:

\[
z_t = \int \varphi_{1s}^*(r - R_i) \hat{H} \varphi_{1s}(r - R_j) \, dr
\]

where \(\varphi_{1s}(r)\) is the 1s orbital wave function, \(\hat{H}\) is the Hamiltonian of the system, and \(R_i\) and \(R_j\) are the positions of the neighboring atoms. In practice, this integral is often approximated using empirical or computational methods, such as density functional theory (DFT).

2. \textbf{On-Site Interaction \(z_u\)}: The on-site interaction \(z_u\) represents the Coulomb repulsion between two electrons occupying the same site. For hydrogen atoms, this can be approximated using the Coulomb integral:

\[
z_u = \int \varphi_{1s}^*(r_1) \varphi_{1s}^*(r_2) \frac{e^2}{|r_1 - r_2|} \varphi_{1s}(r_1) \varphi_{1s}(r_2) \, dr_1 \, dr_2
\]

where \(e\) is the electron charge, and \(\varphi_{1s}(r)\) is the 1s orbital wave function. This integral can also be evaluated using computational techniques, providing an estimate of the electron repulsion energy at each site.